# Chemical hydrodynamics of nuclear spin states

*Anupama Acharya[1], Madhukar Said[2], Sylwia J. Barker[1,3],*
*Marcel Utz[1,3], Bruno Linclau[1,2], and Ilya Kuprov[4,1,*]*

[1]*School of Chemistry and Chemical Engineering, University of Southampton, University Road, Southampton, SO17 1BJ, United Kingdom*

[2]*Department of Organic and Macromolecular Chemistry, Ghent University, Krijgslaan 281-S4, 9000 Ghent, Belgium*

[3]*Institute of Microstructure Technology, Karlsruhe Institute of Technology, Hermann-von-Helmholtz-Platz 1, Eggenstein-Leopoldshafen, 76344, Germany*

[4]*Department of Chemical and Biological Physics, Weizmann Institute of Science, 234 Herzl Street, Rehovot, 76100, Israel*

## Abstract

Quantum mechanical equations of motion are strictly linear in state descriptors, such as wavefunctions and density matrices, but equations describing chemical kinetics and hydrodynamics may be non-linear in concentrations. This incompatibility is fundamental, but special cases can be handled – for example, in magnetic resonance where nuclear spin interactions may be too weak influence concentration dynamics. For processes involving single spins and first-order chemical reactions, this is a well-researched topic, but time evolution of complex nuclear spin systems in the presence of second-order kinetics, diffusion, and flow has so far remained intractable. This creates obstacles in microfluidics, homogeneous catalysis, and magnetic resonance imaging of metabolic processes.

In this communication we report a numerically stable formalism for time-domain quantum mechanical description of nuclear spin dynamics and decoherence in the simultaneous presence of diffusion, flow, and second-order chemical reactions. The formalism is implemented in versions 2.11 and later of the open-source *Spinach* library. As an illustration, we use Diels-Alder cycloaddition of acrylonitrile to cyclopentadiene, yielding endo- and exo-norbornene carbonitrile, in the presence of diffusion and flow in the detection chamber of a microfluidic NMR probe (a finite element model with thousands of Voronoi cells) with a spatially localised stripline radiofrequency coil.

*\*Email: ilya.kuprov@weizmann.ac.il*

# 1. Introduction

Fundamental equations of motion in quantum mechanics of isolated systems [1] and ensembles [2] are required (by causality and time translation invariance [3]), to be linear with respect to state descriptors, such as wavefunctions and density matrices. However, the law of mass action in chemical kinetics [4] and Navier-Stokes equations in hydrodynamics [5,6] are not fundamental; they are statistical approximations, and therefore at liberty to be non-linear with respect to concentrations.

This incompatibility creates insidious difficulties in theoretical descriptions of systems where quantum processes coexist with chemical kinetics and spatial transport, notably in spin chemistry [7], magnetic resonance imaging (MRI) of complex metabolic [8] and hydrodynamic [9] processes, and – our predicament here – nuclear magnetic resonance (NMR) in microfluidic chips [10-12]. The problem involves a collision of approximations at the interface of classical and quantum physics broadly similar to the measurement paradox [13] – although the concentration and the wavefunction amplitude square are both probability densities, one has a specific measurement outcome but the other does not.

The general case has no solution; here we adopt a simplification from condensed phase NMR and assume that nuclear spin processes are influenced by spatial dynamics and chemistry, but that there is no back action because nuclear spin interaction energies are very small. Processes like CIDNP [14] will unfortunately have to wait – there are raging debates about their magnetokinetics, here we focus on reactions that do not involve unpaired electrons. Another generally valid assumption in NMR of diamagnetic systems is that the electronic structure remains in the ground state, only manifesting through effective parameters of the nuclear spin Hamiltonian [15].

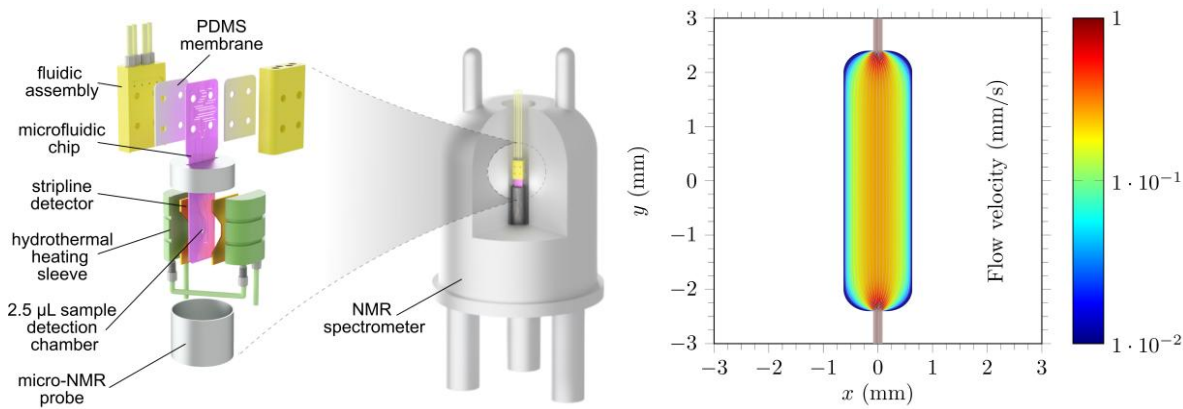

*Figure 1.* Microfluidic NMR experiment as an example of a setting with simultaneous diffusion, hydrodynamics, chemical kinetics, and spatially distributed quantum dynamics in a multi-spin system. *(Left panel)* A schematic diagram of a microfluidic NMR probe [16] with a stripline radiofrequency coil [12]. *(Right panel)* A finite volume simulation, using COMSOL [17], of the velocity field of the stationary fluid flow through the reaction chamber of the chip.

First order kinetics (both exact and approximate) and non-reacting spatial transport of singe spins in magnetic resonance are comprehensively researched and reviewed [18-24]. Simulation of second-order kinetics has been looked at, but the best current formalism is not numerically friendly: concentrations occur in denominators of the equations proposed by Kühne *et al.* [23], meaning that the common case of near-zero concentration yields a singularity that makes those equations numerically unstable. Simultaneous diffusion, flow, second order kinetics, and fully quantum mechanical description of
2

coherent and dissipative spin dynamics in large molecules have not been attempted due to the overwhelming numerical complexity of the task.

At the same time, such simulations are increasingly pertinent – many systems studied by NMR and MRI involve non-linear kinetics. Lactate metabolism is one example: increased lactate levels in mammalian cells ("Warburg effect" [25]) can be an indication of cancer and other disease [26,27]. Increased pyruvate to lactate conversion is a symptom of inflammatory disease in the liver [28] and a sign of injury to the kidneys [29,30], as well as of diabetes [31]. Another example is the tricarboxylic acid cycle: anomalous intermediate concentrations can be an indication of cardiac and neurological disease [32-34]. On the transport side, the simulation of spatially distributed, diffusing, and flowing systems is important in diffusion weighted imaging [35-37], diffusion tensor imaging [38], and vascular imaging by phase-contrast techniques [39-41]. NMR is well integrated with these methods and used for quantification of metabolites [42-44]. However, resolution and strong *J*-coupling problems in proton spectra have caused an exodus towards $^{13}$C and $^{19}$F NMR spectroscopy in metabolomics [45]. Low natural abundance of $^{13}$C also produced a growing emphasis [45-47] on hyperpolarisation techniques such as PHIP [48,49] and DNP [50,51]. Pyruvate is a common target [8,52-54] as a key intermediate and a branching point for further metabolism [34].

No existing simulation software can handle this level of chemical, spatial, and spin dynamics complexity. Major existing packages, such as *SIMPSON* [55], *GAMMA* [56], and *SpinEvolution* [57] treat spin quantum mechanically, but only cover solid orientation distributions. On the MRI side, packages such as *SIMRI* [58], *coreMRI* [59], *JEMRIS* [60], and *MRISMUL* [61] can model sophisticated spatial dynamics, but use Bloch-Torrey equations [62] for spin. Very few MRI packages [63] implement Bloch-McConnell [64] solvers; most implementations are standalone simulation frameworks mainly focusing on chemical exchange saturation transfer MRI [65-67].

In this communication, we report a theoretical formalism and a *Spinach* [68] implementation for the full non-linear kinetics + magnetohydrodynamics case. The problem is a quantum mechanical generalisation of the Fokker-Planck formalism [69,70] in which concentration is replaced by concentration-weighted density matrix [71,72] and the evolution generator is both time- and state-dependent. Its efficient numerical implementation is difficult: dimensions of spatial dynamics generator matrices on finite grids can be large; when combined with the nuclear spin density matrix of a typical metabolite, the composite evolution generators cannot even be stored, let alone manipulated. This problem was recently solved by Allami *et al.* [73]; here we build on their methods by storing the combined evolution generators in a polyadic format with buffered Kronecker products. For logistical reasons also discussed in [73], the state vector remains uncompressed.

We apply the resulting software to microfluidic chip NMR experiments (Figure 1) where we model simultaneous diffusion, flow, and spin dynamics during a second-order cycloaddition reaction. The chemistry and the engineering are described in [16,74]: reactants are delivered at the upper end of the chip, the products are detected *via* their NMR signals at the lower end of the sample chamber. The spatial discretisation mesh and the velocity field are imported from *COMSOL* [17]. At the nuclear spin dynamics simulation level, chemical kinetics is described using state-dependent superoperators [72] acting on density matrices in each Voronoi cell of the mesh. Quantum mechanical treatment of coupled multi-spin systems, essential in such processes, is therefore maintained.

## 2. Equations of motion

In this section we build a numerically friendly equation of motion for the concentration-weighted density matrix $\boldsymbol{\eta} = c\boldsymbol{\rho}$ under the assumption that kinetics, diffusion, and flow do not depend on the nuclear spin state. Chemical concentration $c$ is a type of probability density and eigenvalues of the density matrix $\boldsymbol{\rho}$ are probabilities; eigenvalues of $\boldsymbol{\eta}$ are therefore also probability densities. The density matrix must be thermodynamically correct (*i.e.* the zero-trace tomfoolery [75,76] is not permitted) to reflect the fact that only a fraction of each substance is spin-polarised; then $c_n = \mathrm{Tr}(\boldsymbol{\eta}_n)$, where the index $n$ runs over chemical substances. The individual processes (kinetics, diffusion and flow, spin dynamics) are described by Lie semigroup actions on the corresponding state spaces; we then merge their algebras [72,77] to bring them together.

### 2.1 Chemical kinetics

Within the assumptions made by the law of mass action [4], a network of elementary chemical reactions involving $N$ substances obeys the following equations:

$$\frac{\partial c_n}{\partial t} = f_n(c_1,\ldots,c_N) \tag{1}$$

where $c_n$ is the concentration of *n*-th substance, $f_n$ are low-order polynomials, and the partial derivative is a reminder that these equations govern *local* kinetics at each point of a three-dimensional sample. We assume that nuclear spin state has no effect on this dynamics.

### 2.2 Diffusion and flow

We assume that Fick's first Law [78] is also unaffected by the nuclear spin state:

$$\mathbf{j}_n = \mathbf{v} c_n - D_n \nabla c_n \tag{2}$$

where $\mathbf{j}_n$ is the net flux of the *n*-th substance, $D_n$ is its diffusion coefficient, $\mathbf{v}$ is the velocity of the fluid flow (may be time-dependent), and $\nabla$ is the gradient operator with respect to the location within the sample. The continuity equation [79] then relates flux divergence to concentration change, and the kinetic terms are inherited from Eq (1):

$$\frac{\partial c_n}{\partial t} = -\nabla \cdot \mathbf{j}_n + f_n(c_1,\ldots,c_N) \tag{3}$$

Up to this point, everything is standard [80]; this equation may be solved independently. We therefore assume below that concentrations had been pre-computed as functions of time and location; the particular numerical implementations used by *Spinach* are described in Section 3 below.

Very strictly speaking, thermodynamically correct diffusive flux in Eq (2) should use gradients of thermodynamic activities, not concentrations. That is not a problem – activity coefficients may be packed into an effective concentration-dependent diffusion constant. The velocity flux and the conservation law embodied in Eq (3) do strictly refer to concentrations – we can therefore still proceed.

## 2.3 Chemical transport of nuclear spin state

Concentrations can be zero or near-zero. For this reason, the formalism published by Kühne *et al.* [23] is unstable in finite precision arithmetic – their Eqs 10 and 11 have concentrations in denominators. This was not a concern in 1979 when analytical solutions were the preferred way forward, but here we must build a formalism in which the computer never divides by a concentration, and preferably never divides at all because the inverse might not exist under semigroup dynamics.

We must also account for the fact that spin state spaces may be different on either side of the reaction arrow; this is illustrated in Figure 2. For example, the association process:

$$A + B \rightleftarrows C \qquad (4)$$

maps a direct sum of state spaces of reactants A and B (in a solution, reactant molecules would normally be uncorrelated in their nuclear spin state) into a direct product of those spaces that describes the substance C. The map induced by the forward reaction is clearly 1-to-1, but the reverse reaction is a dissociation process – some nuclear spin correlations in substance C would be broken up and lost because those two particular shards A and B are unlikely to meet again in a typical solution. Thorny complications that can arise here are debated in the spin chemistry literature [81-83]; those are only significant for electron spin dynamics, here we deal with nuclei.

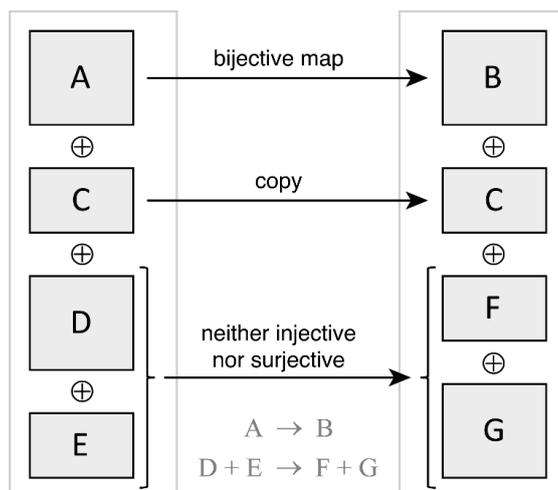

*Figure 2. Maps between nuclear spin density matrix spaces under three types of chemical processes in a system containing substances {A,C,D,E} that are chemically converted into substances {B,C,F,G} as shown in the figure. Isomerisation (A → B) changes the Hamiltonian, but does not change the density matrix space; this is the well-researched case of a first-order chemical reaction going back to McConnell [64]. Observer substances, for example solvent, simply continue evolving under the same Hamiltonian (C → C). This paper resolves the implementation difficulties associated with bimolecular reactions such as (D + E → F + G) where the nuclear spin density matrix space on either side of the reaction arrow need not be the same: some product states may be lost, some moved over, and some rearranged.*

We assume that a nucleus, when moving from a reactant to a product, takes its spin state and ensemble correlations with it. In complete Liouville spaces that have a direct product structure [84], this assumption leads directly to the kinetics superoperator [19]. However, when an incomplete Liouville space is used to treat a large spin system [85] the procedure is more nuanced: we must consider individually the structure and the population of each state in the incomplete basis and account for the initial location and the destination of each nucleus. For each chemical reaction, the kinetic part of the equation of motion is therefore built as described in [72]. The algorithm is:

1. Index the Liouville space basis $\{\boldsymbol{\beta}_p^{(n)}\}$ of nuclear spin states for every substance involved in the reaction. Here $n$ runs over substances and $p$ over basis states. Each basis state $\boldsymbol{\beta}_p^{(n)}$ is a direct product of single-nucleus spin operators of the form:

$$\mathbf{T}_{l_1,m_1} \otimes \mathbf{T}_{l_2,m_2} \otimes \mathbf{T}_{l_3,m_3} \otimes \cdots \Leftrightarrow \begin{pmatrix} l_1 & l_2 & l_3 & \cdots \\ m_1 & m_2 & m_3 & \cdots \end{pmatrix} \quad (5)$$

where $l, m$ are state indices of the single-spin irreducible spherical tensor basis set [15]. Only the indices need to be stored because they define the operator unambiguously. For large spin systems, this basis may be incomplete [85].

2. Using the indexed representation in Eq (5), build a matching table of nuclear spin basis states on either side of the reaction arrow, indicating which state on the left of the reaction arrow is mapped into which state on the right. The following is possible:

   (a) an identical state involving the same spins exists in the destination basis – the population of the state is then to be drained from the source state space and replenished in the destination state space;

   (b) an identical state involving the same spins does not exist in the destination basis, for example because it becomes intermolecular – its population is then to be drained at the source but not forwarded to any destination.

   The result is a set of superoperators which we call *drain generators* $\mathcal{D}_{rs}$ (sparse matrices with −1 on the diagonal for each nuclear spin state that is drained by reaction $r$ from substance $s$) and *fill generators* $\mathcal{F}_{rs}$ (sparse matrices with +1 between the source and the destination state whenever the destination exists in the basis set). These matrices are evolution generators in the Lie semigroup sense – to produce propagators acting on the system state vector, they need to be multiplied by the corresponding rates, added up, and exponentiated with the chosen time step.

3. At each point of the sample, the local concentration-weighted density matrix $\boldsymbol{\eta}_n$ of each substance $n$ then evolves according to the following equation:

$$\frac{\partial}{\partial t}\begin{pmatrix} \boldsymbol{\eta}_1 \\ \boldsymbol{\eta}_2 \\ \vdots \end{pmatrix} = -i \begin{pmatrix} \mathcal{H}_1(\mathbf{r},t)\boldsymbol{\eta}_1 \\ \mathcal{H}_2(\mathbf{r},t)\boldsymbol{\eta}_2 \\ \vdots \end{pmatrix} + \begin{pmatrix} \mathcal{R}_1(\mathbf{r},t)\boldsymbol{\eta}_1 \\ \mathcal{R}_2(\mathbf{r},t)\boldsymbol{\eta}_2 \\ \vdots \end{pmatrix} + \\ + \sum_r k_r \sum_{s \in r} \left(\prod_{m \neq s} c_m\right)(\mathcal{D}_{rs} + \mathcal{F}_{rs})\begin{pmatrix} \boldsymbol{\eta}_1 \\ \boldsymbol{\eta}_2 \\ \vdots \end{pmatrix}, \quad c_n = \mathrm{Tr}(\boldsymbol{\eta}_n) \quad (6)$$

where Hamiltonian commutation superoperators $\mathcal{H}_n$ may be time- and location-dependent. Relaxation superoperators $\mathcal{R}_n$ may be location-dependent (for example, due to variations in local viscosity [86], magnetic field, or other parameters) but are not usually intrinsically time-dependent. In the kinetics superoperator, the outer sum is over the chemical reactions, $k_r$ are their rate constants from the mass action law. The $s$ index of the inner sum enumerates reactants; each of them forwards its spin states to the products as prescribed by the drain generator $\mathcal{D}_{rs}$ and the fill generator $\mathcal{F}_{rs}$. The rate of that process is proportional to the concentration product, but reactant's own concentration is already present in the concentration-weighted density matrix – therefore only the remaining concentrations are multiplied up.

At each spatial location, Eq (6) describes nuclear spin dynamics driven by coherent evolution, relaxation, and chemical redistribution of spin state populations. It holds simultaneously with Eq (3) that governs spatial dynamics and chemical kinetics. They are connected by the concentration dependence of the chemical transport rates and the fact that the physical quantity seen by the NMR instrument is the concentration-weighted density matrix.

## 2.4 Spatial transport of nuclear spin state

When substances with non-equilibrium nuclear spin states are introduced into the system, they are moved around by diffusion and flow. Here we also assume that spatial transport does not depend on the nuclear spin state; this is exceedingly well studied in the Bloch equation limit [87-89]. For the density matrix, the equation enforcing the conservation of probability reads:

$$\frac{\partial \boldsymbol{\eta}_n}{\partial t} = -\nabla \cdot \left( \mathbf{v} \otimes \boldsymbol{\eta}_n - D_n \nabla \boldsymbol{\eta}_n \right) \tag{7}$$

where two terms in the brackets are fluxes of $\boldsymbol{\eta}$ due to flow and diffusion. It reflects the fact that concentration gradients are not required for the density matrix transport to occur – spin states still move around even if concentrations are uniform. Importantly, Eq (7) does not require us to keep track of concentration and spin state separately; only their product is needed – a significant logistical improvement over the state of the art [23].

## 2.5 Combined equations of motion

With individual components now in place, we must solve two systems of partial differential equations. Firstly, the nuclear spin independent diffusion, flow, and chemical kinetics:

$$\frac{\partial c_n}{\partial t} = -\nabla \cdot \left[ \mathbf{v}(\mathbf{r},t) c_n - D_n \nabla c_n \right] + f_n(c_1,\ldots,c_N) \tag{8}$$

where initial and boundary conditions are specified by the user. This stage is well researched [80]; after solving Eq (8), we get time and location dependence of all concentrations $c_n(\mathbf{r},t)$.

The second system of equations is the balance of probability for the concentration-weighted density matrix of each substance, including all evolution generators discussed above:

$$\frac{\partial}{\partial t}\begin{pmatrix} \boldsymbol{\eta}_1 \\ \boldsymbol{\eta}_2 \\ \vdots \end{pmatrix} = -\nabla \cdot \begin{pmatrix} \mathbf{v} \otimes \boldsymbol{\eta}_1 - D_1 \nabla \boldsymbol{\eta}_1 \\ \mathbf{v} \otimes \boldsymbol{\eta}_2 - D_2 \nabla \boldsymbol{\eta}_2 \\ \vdots \end{pmatrix} - i \begin{pmatrix} \mathcal{H}_1(\mathbf{r},t) \boldsymbol{\eta}_1 \\ \mathcal{H}_2(\mathbf{r},t) \boldsymbol{\eta}_2 \\ \vdots \end{pmatrix} \\ + \begin{pmatrix} \mathcal{R}_1(\mathbf{r},t) \boldsymbol{\eta}_1 \\ \mathcal{R}_2(\mathbf{r},t) \boldsymbol{\eta}_2 \\ \vdots \end{pmatrix} + \sum_r k_r \sum_{s \in r} \left( \prod_{m \neq s} c_m \right) (\mathcal{D}_{rs} + \mathcal{F}_{rs}) \begin{pmatrix} \boldsymbol{\eta}_1 \\ \boldsymbol{\eta}_2 \\ \vdots \end{pmatrix}, \quad c_n = \mathrm{Tr}(\boldsymbol{\eta}_n) \tag{9}$$

where elements of column vectors refer to different substances. Spin evolution generators can act on $\boldsymbol{\eta}$ instead of $\boldsymbol{\rho}$ because they commute with the scalar concentration multiplier.

Relaxation superoperators $\mathcal{R}_n$ in Eq (9) are rarely explicitly time-dependent but commonly location-dependent; that is accounted for by assigning a different relaxation superoperator to each cell of the mesh. A more subtle matter is that the thermal equilibrium state is concentration-dependent:

$$\boldsymbol{\eta}_n^{(eq)} = c_n \exp(-\mathbf{H}_n/kT) \big/ \mathrm{Tr}\big[\exp(-\mathbf{H}_n/kT)\big] \tag{10}$$

In the conventional formulation of the Liouville space (as the adjoint representation of the Hilbert space [84]), this would require the relaxation superoperator to be updated every time concentration changes – an inconvenient and numerically expensive process. However, the indexed product state basis sets used by *Spinach* (described in detail in [85]) have the trace of the density matrix as the first element of the state vector. When concentration weighting is then performed, the first element of the state vector ends up being concentration, meaning that concentration-dependent thermalisation within the inhomogeneous master equation (IME) formalism [90] takes care of itself:

$$\begin{aligned}\frac{\partial \boldsymbol{\rho}}{\partial t} &= \ldots + \mathcal{R}(\boldsymbol{\rho} - \boldsymbol{\rho}_{eq}) & \Rightarrow & \quad \frac{d}{dt}\begin{pmatrix}1\\\boldsymbol{\rho}\end{pmatrix} = \ldots + \begin{bmatrix}0 & 0\\-\mathcal{R}\boldsymbol{\rho}_{eq} & \mathcal{R}\end{bmatrix}\begin{pmatrix}1\\\boldsymbol{\rho}\end{pmatrix}\\\frac{\partial c\boldsymbol{\rho}}{\partial t} &= \ldots + \mathcal{R}(c\boldsymbol{\rho} - c\boldsymbol{\rho}_{eq}) & \Rightarrow & \quad \frac{d}{dt}\begin{pmatrix}c\\c\boldsymbol{\rho}\end{pmatrix} = \ldots + \begin{bmatrix}0 & 0\\-\mathcal{R}\boldsymbol{\rho}_{eq} & \mathcal{R}\end{bmatrix}\begin{pmatrix}c\\c\boldsymbol{\rho}\end{pmatrix}\end{aligned} \tag{11}$$

Note that the operator in square brackets is the same in both lines; this automatic concentration scaling behaviour is a welcome feature – it means that each relaxation superoperator needs only to be thermalised once to drive the system to:

$$\boldsymbol{\rho}_n^{(eq)} = \exp(-\mathbf{H}/kT) \big/ \mathrm{Tr}\big[\exp(-\mathbf{H}/kT)\big] \tag{12}$$

As required, and unlike the formalism presented in [23], there are no denominators in Eq (9); it is therefore stable in finite precision arithmetic when concentrations are close to zero. It is solved using standard numerical methods discussed in Section 3; we recommend Lie quadratures that respect group-theoretical constraints [91-94], they were also recently implemented in *Spinach* [95].

## 3. Implementation details

Coding up a solver and visualisation tools for Eqs (8) and (9) is a harder exercise than the straightforward mathematical derivation we just gave it. Omitting software implementation details is something theory papers are too often guilty of; we break with that tradition here.

### 3.1 Mesh and flow velocity data handling

*Matlab* does have a meshing tool for arbitrary domains [96], but in our context it was more convenient to import the mesh from COMSOL [17] because that is where the stationary flow velocity field had been computed by the experimental team. ASCII files containing mesh information were parsed and mesh elements classified into edges, triangles, and rectangles; the latter are used by COMSOL to indicate impenetrable walls (Figure 3, left).

*Spinach* stores mesh information as a data structure (Figure 3, right) with the fields containing vertex coordinates and three index arrays: pairs of integers indicating which vertices make an edge, triads of integers for triangles, and tetrads of integers for rectangles. Voronoi tessellation information, computed by *Matlab*, is added as further arrays with coordinates of Voronoi vertices, integer indices for which vertices belong to which Voronoi cell, and weights indicating the volume of each cell.

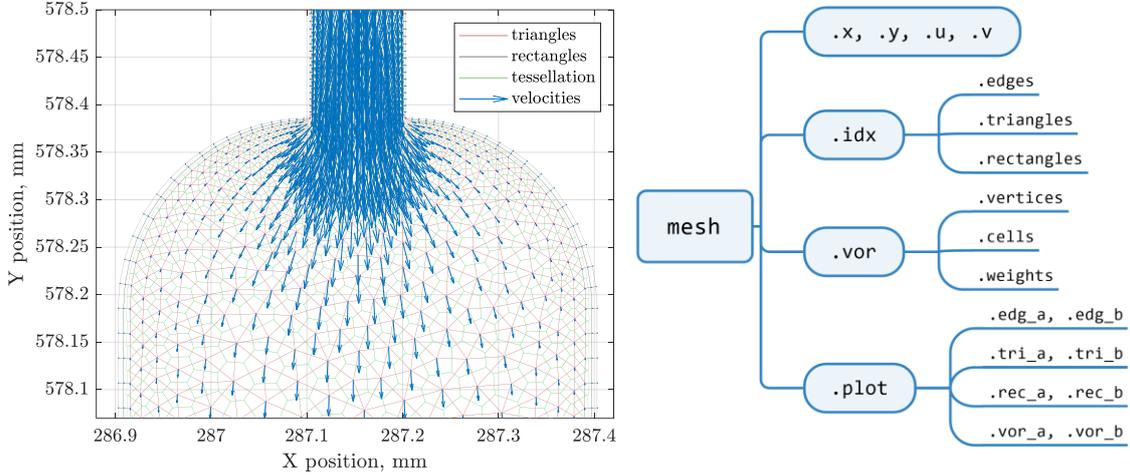

*Figure 3. (Left)* A numerical model of the sample chamber of the microfluidic chip described in [16] with an adaptive grid (red lines) produced by COMSOL [17]. Blue arrows indicate stationary flow velocities and green lines are Voronoi tessellation produced by Matlab [96]. Solutions of the flow and the chemistry problems are handled by COMSOL; stationary concentration and flux fields are then exported into Spinach 2.10 [68] which handles the nuclear spin dynamics as described in the main text. *(Right)* Spinach 2D mesh data structure schematic, including coordinate and velocity arrays, mesh and tessellation indices, and pre-computed information to facilitate the plotting as discussed in the main text.

On the visualisation side, *Matlab* figures are objects – when a new element is added, a new sub-object is created. It is therefore impractical to draw complex meshes line by line – the list of sub-objects becomes too large for interactive plotting. For this reason, further arrays are pre-computed, containing Cartesian coordinates of the endpoints of each line separated by NaN values. When *Matlab* encounters NaN values in coordinate arrays, it creates line breaks; this allows the entire mesh to be plotted as one object which accelerates interactive graphics. All of the above arrays are stored in the `.mesh` subfield of the *Spinach* `spin_system` data structure (Figure 3, right).

### 3.2 Discretisation of the equation of motion

A gridded domain, such as that in Figure 3, has a set of locations corresponding to Voronoi cells numbered by the index $k$ with centres at $\mathbf{r}_k$. Each location has a set of substance concentrations $c_{kn}$ (first index refers to location, second to the substance) and a set of flux vectors $\mathbf{j}_{kn}$. Each substance at each location has a nuclear spin density matrix $\boldsymbol{\rho}_{kn}$ with a unit trace. The objective is to calculate the dynamics of concentration-weighted density matrices $\boldsymbol{\eta}_{kn} = c_{kn} \boldsymbol{\rho}_{kn}$.

We import the (possibly time-dependent) flow velocity field from specialised software, in this case *COMSOL* [17]. Without the reaction terms already discussed above, the continuous forms of Eqs (2) and (3) for each substance in a stationary flow and diffusion regime are:

$$\partial c(\mathbf{r},t)/\partial t = -\nabla \cdot \mathbf{j}(\mathbf{r},t) \tag{13}$$

$$\mathbf{j}(\mathbf{r},t) = \mathbf{v}(\mathbf{r})c(\mathbf{r},t) - D\nabla c(\mathbf{r},t) \tag{14}$$

where the flux vector $\mathbf{j}(\mathbf{r},t)$ has an advection component from the flow velocity field $\mathbf{v}(\mathbf{r})$ and a diffusion component proportional to the concentration gradient.

The standard finite volume algorithm (a schematic for the 2D case is shown in Figure 4), is essentially a set of conservation laws that balance local concentration changes with boundary integrals of fluxes [97]. For didactic purposes, we include here a derivation for the 2D case; the three-dimensional case may be found in the hydrodynamics literature [98,99].

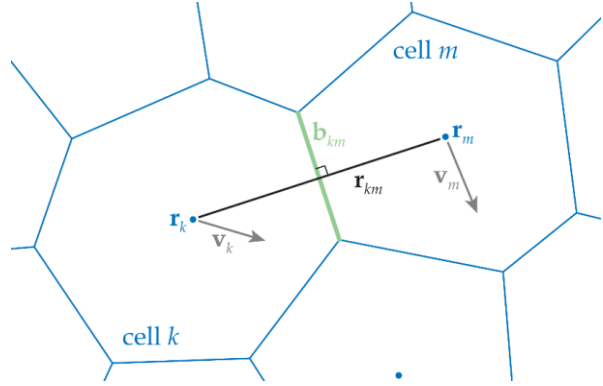

*Figure 4.* Finite volume hydrodynamics solver setup schematic in two dimensions. Blue lines are edges of Voronoi tessellation cells, blue dots are cell centres, grey arrows are velocity vectors. The local rate of change in substance concentration in each cell is computed as a balance of advective and diffusive fluxes through cell boundaries.

We start by integrating Eq (13) over the area $A_k$ of the Voronoi cell $k$ and note that the time derivative on the left hand side commutes with area integration:

$$\frac{\partial}{\partial t}\iint_{A_k} c \, dA = -\iint_{A_k} \nabla \cdot \mathbf{j} \, dA \tag{15}$$

After we apply the divergence theorem [100] on the right hand side, this becomes:

$$\frac{\partial}{\partial t}\iint_{A_k} c \, dA = -\oint_{B_k} \mathbf{j} \cdot \mathbf{n} \, dB \tag{16}$$

where $B_k$ is the boundary of the Voronoi cell $k$ and $\mathbf{n}$ is the outward normal vector. On the left hand side we now assume that the mesh is fine enough for the concentration $c_k$ to be constant within each cell $k$. On the right-hand side, we break the contour integral up into integrals along each boundary segment of the cell (Figure 4 illustrates the notation):

$$A_k \frac{\partial c_k}{\partial t} = -\sum_{m \in \mathcal{N}_k} \int_{B_{km}} \mathbf{j} \cdot \mathbf{n}_{km} \, dB, \quad \mathbf{n}_{km} = \frac{\mathbf{r}_{km}}{|\mathbf{r}_{km}|} \tag{17}$$

where $\mathcal{N}_k$ is the set of neighbouring cells to cell $k$, $B_{km}$ is the shared boundary of cells $k$ and $m$, and the expression for $\mathbf{n}_{km}$ follows from the definition of Voronoi tessellation [101,102]. We now proceed to use Eq (14) and approximate flux integrals through each boundary as:

$$\int_{B_{km}} (\mathbf{v}c) \cdot \mathbf{n}_{km} \, dB \approx |\mathbf{b}_{km}| \frac{\mathbf{v}_k c_k + \mathbf{v}_m c_m}{2} \cdot \frac{\mathbf{r}_{km}}{|\mathbf{r}_{km}|} \tag{18}$$

$$\int_{B_{km}} (D\nabla c) \cdot \mathbf{n}_{km} dB \approx D|\mathbf{b}_{km}| \frac{c_m - c_k}{|\mathbf{r}_{km}|} \tag{19}$$

where $|\mathbf{b}_{km}|$ is the length of the shared boundary of cells $k$ and $m$, and the flux is the average flux at the boundary of the two neighbouring cells. After cosmetic rearrangements:

$$\frac{\partial c_k}{\partial t} = \sum_{m \in \mathcal{N}_k} \frac{D|\mathbf{b}_{km}|}{A_k} \frac{c_m - c_k}{|\mathbf{r}_{km}|} - \sum_{m \in \mathcal{N}_k} \frac{|\mathbf{b}_{km}|}{A_k} \frac{\mathbf{v}_k c_k + \mathbf{v}_m c_m}{2} \cdot \frac{\mathbf{r}_{km}}{|\mathbf{r}_{km}|} \tag{20}$$

In a matrix representation, for the spatial transport generator $\mathbf{F}$ acting on a column vector $\mathbf{c}$ of substance concentrations in each Voronoi cell:

$$\frac{d\mathbf{c}}{dt} = \mathbf{F}\mathbf{c}, \quad F_{km} = \begin{cases} \frac{1}{A_k} \frac{|\mathbf{b}_{km}|}{|\mathbf{r}_{km}|} \left( D - \frac{\mathbf{v}_m \cdot \mathbf{r}_{km}}{2} \right) & m \in \mathcal{N}_k \\ 0 & m \notin \mathcal{N}_k \end{cases}, \quad F_{kk} = -\sum_{m \neq k} F_{mk} \tag{21}$$

where $F_{kk}$ are computed using the conservation of matter balance because it must be enforced to machine precision. Examples of concentration evolution under this equation are shown in Section 4.2 below; annotated *Matlab* code is released as a part of the open-source *Spinach* library [68].

When the flux generator $\mathbf{F}$ acts on the concentration-weighted density matrix $\boldsymbol{\eta}$, it should be extended to a Kronecker product $\mathbf{F} \otimes \mathbf{1}$ that acts with a unit matrix on the spin subspace because the dynamics it generates is nuclear spin independent. With that in place, continuous degrees of freedom are now discretised, and Eq (9) acquires a pure matrix-vector form:

$$\frac{\partial}{\partial t} \begin{pmatrix} \boldsymbol{\eta}_1 \\ \boldsymbol{\eta}_2 \\ \vdots \end{pmatrix} = \begin{pmatrix} (\mathbf{F} \otimes \mathbf{1})\boldsymbol{\eta}_1 \\ (\mathbf{F} \otimes \mathbf{1})\boldsymbol{\eta}_2 \\ \vdots \end{pmatrix} - i \begin{pmatrix} \mathcal{H}_1(\mathbf{r},t)\boldsymbol{\eta}_1 \\ \mathcal{H}_2(\mathbf{r},t)\boldsymbol{\eta}_2 \\ \vdots \end{pmatrix} + \\ + \begin{pmatrix} \mathcal{R}_1(\mathbf{r},t)\boldsymbol{\eta}_1 \\ \mathcal{R}_2(\mathbf{r},t)\boldsymbol{\eta}_2 \\ \vdots \end{pmatrix} + \sum_r k_r \sum_{s \in r} \left( \prod_{m \neq s} c_m(\mathbf{r},t) \right) (\mathcal{D}_{rs} + \mathcal{F}_{rs}) \begin{pmatrix} \boldsymbol{\eta}_1 \\ \boldsymbol{\eta}_2 \\ \vdots \end{pmatrix} \tag{22}$$

where diffusion, flow, spin evolution and relaxation happen independently for each chemical substance. The corresponding blocks of the equation of motion are linked by chemical transport of nuclear spin states with concentration-dependent rates.

### 3.3 Chemical reaction specification

Spin system specification in *Spinach* is described in detail in the program documentation; simple tutorials are also available in our recent paper [103]; here we focus on the new syntax associated with the non-linear kinetics and on the way chemical reaction generators $\mathcal{D}_{rs}$ and $\mathcal{F}_{rs}$ in Eq (22) are built in the actual code. The first logistical hurdle is the need to re-index interaction arrays when multiple non-interacting spin systems are brought together into a single data structure.

To that end, we have implemented a new `merge_inp` function in *Spinach* 2.10 that combines and re-indexes multiple `sys` and `inter` input structures [103] automatically:

```
[sys,inter]=merge_inp({sys_a,sys_b},{inter_a,inter_b});
```

and also extends the rotational correlation time variable to an array that holds one rotational correlation time (a scalar or a tensor when rotational diffusion is anisotropic) per chemical species. That may be necessary even when molecules are not affected by a reaction – for example, when viscosity is different on either side of the cellular membrane in a translocation process [86].

Chemical species indexing in *Spinach* is unchanged since the description given in [103]. Consider a simple cycloaddition reaction involving $^{12}$C carbons so that only protons have spin:

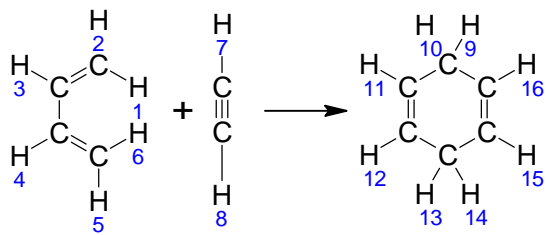

for this reaction, the part specification is a cell array of integer sequences:

```
inter.chem.parts={1:6, 7:8, 9:16};
```

It indicates that there are three chemical species, the first one contains spins 1 to 6, the second one spins 7 and 8, and the third one spins from 9 to 16. These protons and their corresponding interactions are expected to have been declared earlier in the input stream as described in [103].

We now come to the new input structure in *Spinach* 2.10, called `reaction`, that supports non-linear kinetics. Its `reactants` subfield is an array of integers specifying which chemical species are on the left side of the reaction arrow; the `products` subfield does the same for the chemical species on the right side of the reaction arrow. For the cycloaddition reaction above:

```
reaction.reactants=[1 2];
reaction.products=[3];
```

The information about which spin in the reactant set becomes which spin in the product set is provided as a matching table with pairs of integers. For the cycloaddition reaction above:

```
reaction.matching=[1 9; 2 10; 3 11; 4 12; 5 13; 6 14; 7 16; 8 15];
```

Combined with the basis set specification built as described in [85], this allows *Spinach* to build reaction drain and fill generators as described in Section 2.3 above in a single user command:

```
G=react_gen(spin_system,reaction);
```

When multiple reactions are present in the system, this function is called multiple times with different `reaction` specifications to obtain the corresponding matrices $\mathcal{D}_{rs}$ and $\mathcal{F}_{rs}$ which are returned, as elements of the cell array `G`, by `react_gen.m` function. As discussed in Section 2.3, they are drain and fill generators in the Lie algebraic sense: they do not depend on concentrations or time – they only indicate what goes where in the spin state space of the system as a result of each reaction. Concentration dependence turns up in their coefficients in Eq (9).

A subtle point is that the cycloaddition reaction above yields not one but four products: one for acetylene binding from the top, one from the bottom, and two across. In the absence of selective isotope labelling, such topological matters are inconsequential to a chemist, but they may be significant to an NMR spectroscopist in the rare cases where the reactants have non-symmetric nuclear spin states. In those cases, multiple reactions with different matching tables must be specified.

### 3.4 Generator and state vector structure

For logistical reasons (to do with object layouts in *Matlab*), *Spinach* uses the following order of direct products in evolution generators that have spatial, chemical, and spin degrees of freedom [73]:

$$\mathbf{G}(t) = \sum_{nmk} g_{nmk}(t)\, \mathbf{M}_n \otimes \mathbf{K}_m \otimes \mathbf{S}_k \qquad (23)$$

where $g_{nmk}(t)$ are interaction coefficients, $\mathbf{M}_n$ are spatial dynamics generators, $\mathbf{K}_m$ are chemical kinetics generators, and $\mathbf{S}_k$ are spin dynamics generators. Accordingly, the initial state vector $\boldsymbol{\eta}(t_0)$ is built from concentrations and spin states of each substance in the following way:

$$\boldsymbol{\eta}(t_0) = \begin{bmatrix} \begin{pmatrix} c_1^{(1)}(t_0) \\ \vdots \\ c_N^{(1)}(t_0) \end{pmatrix} \odot \begin{pmatrix} \boldsymbol{\rho}_1^{(1)}(t_0) \\ \vdots \\ \boldsymbol{\rho}_N^{(1)}(t_0) \end{pmatrix} \\ \vdots \\ \begin{pmatrix} c_1^{(K)}(t_0) \\ \vdots \\ c_N^{(K)}(t_0) \end{pmatrix} \odot \begin{pmatrix} \boldsymbol{\rho}_1^{(K)}(t_0) \\ \vdots \\ \boldsymbol{\rho}_N^{(K)}(t_0) \end{pmatrix} \end{bmatrix} \qquad (24)$$

where $\odot$ denotes element-wise multiplication, $c_n^{(k)}$ is the concentration of the *n*-th substance in the *k*-th Voronoi cell of the mesh and $\boldsymbol{\rho}_n^{(k)}$ is the corresponding nuclear spin density matrix represented by a vector in a full or restricted Liouville space. There are no multiplicative ambiguities here because density matrices have unit traces.

When the state vector is evolved in time under Eq (22), concentrations and density matrices can no longer be separated without loss of numerical stability, and therefore only products $\boldsymbol{\eta}_n^{(k)} = c_n^{(k)} \boldsymbol{\rho}_n^{(k)}$ are stored in the trajectory array $\mathbf{T}$ at each time discretisation point $t_m$:

$$\mathbf{T} = \begin{bmatrix} \begin{pmatrix} \boldsymbol{\eta}_1^{(1)}(t_0) \\ \vdots \\ \boldsymbol{\eta}_N^{(1)}(t_0) \end{pmatrix} & \begin{pmatrix} \boldsymbol{\eta}_1^{(1)}(t_1) \\ \vdots \\ \boldsymbol{\eta}_N^{(1)}(t_1) \end{pmatrix} & \begin{pmatrix} \boldsymbol{\eta}_1^{(1)}(t_2) \\ \vdots \\ \boldsymbol{\eta}_N^{(1)}(t_2) \end{pmatrix} & \cdots \\ \vdots & \vdots & \vdots & \cdots \\ \begin{pmatrix} \boldsymbol{\eta}_1^{(K)}(t_0) \\ \vdots \\ \boldsymbol{\eta}_N^{(K)}(t_0) \end{pmatrix} & \begin{pmatrix} \boldsymbol{\eta}_1^{(K)}(t_1) \\ \vdots \\ \boldsymbol{\eta}_N^{(K)}(t_1) \end{pmatrix} & \begin{pmatrix} \boldsymbol{\eta}_1^{(K)}(t_2) \\ \vdots \\ \boldsymbol{\eta}_N^{(K)}(t_2) \end{pmatrix} & \cdots \end{bmatrix} \qquad (25)$$

This array may be stored explicitly or (for efficiency reasons) only the previous state vector may be kept at each point in the time-domain simulation.

The detection state $\boldsymbol{\delta}$ is built in the same way as the initial condition in Eq (24), but the role of concentration is played by the $B_+$ map of each radiofrequency coil. It may be different in different parts of the sample because the coil field vector may be different:

$$\boldsymbol{\delta} = \begin{pmatrix} b_+^{(1)} \\ \vdots \\ b_+^{(K)} \end{pmatrix} \otimes \begin{pmatrix} \mathbf{L}_+^{(1)} \\ \vdots \\ \mathbf{L}_+^{(N)} \end{pmatrix} = \begin{bmatrix} \begin{pmatrix} b_+^{(1)}\mathbf{L}_+^{(1)} \\ \vdots \\ b_+^{(1)}\mathbf{L}_+^{(N)} \end{pmatrix} \\ \vdots \\ \begin{pmatrix} b_+^{(K)}\mathbf{L}_+^{(1)} \\ \vdots \\ b_+^{(K)}\mathbf{L}_+^{(N)} \end{pmatrix} \end{bmatrix} \quad (26)$$

where $b_+^{(k)}$ is the receptivity of the coil in the *k*-th Voronoi cell of the mesh and $\mathbf{L}_+^{(n)} = \mathbf{L}_X^{(n)} + i\mathbf{L}_Y^{(n)}$ is the Liouville space representation of the quadrature detection operator of the *n*-th substance. Taking the inner product of $\boldsymbol{\delta}$ with the trajectory array (but note the absence of complex conjugation in $O = \mathrm{Tr}[\mathbf{O}\boldsymbol{\rho}]$) yields a quantity proportional to the voltage induced in the detection coil:

$$\langle\boldsymbol{\delta}|\mathbf{T} = \sum_{kn} b_+^{(k)} \begin{bmatrix} \mathrm{Tr}\left(\mathbf{L}_+^{(n)}\boldsymbol{\eta}_n^{(k)}(t_0)\right) & \mathrm{Tr}\left(\mathbf{L}_+^{(n)}\boldsymbol{\eta}_n^{(k)}(t_1)\right) & \mathrm{Tr}\left(\mathbf{L}_+^{(n)}\boldsymbol{\eta}_n^{(k)}(t_2)\right) & \cdots \end{bmatrix} \quad (27)$$

in which the observable is weighted with both the concentration (through the use of concentration-weighted density matrix) and the coil receptivity (through the use of the coil map).

Block structure of evolution generators matches the state vector layout described above, taking into account location and concentration dependence of the individual terms. Hamiltonian and relaxation superoperators may be different for each substance $n$ in each Voronoi cell $k$; they are therefore built from "phantoms" pertaining to individual interactions and relaxation mechanisms.

$$\mathcal{H}_n^{(k)} = \sum_j \varphi_{nkj}^{(\mathrm{H})} \mathcal{B}_{nj}^{(\mathrm{H})}, \qquad \mathcal{R}_n^{(k)} = \sum_j \varphi_{nkj}^{(\mathrm{R})} \mathcal{B}_{nj}^{(\mathrm{R})} \quad (28)$$

where the sums run over spin interactions and relaxation mechanisms, $\mathcal{B}_{nj}^{(\mathrm{H})}$ is a basis set of superoperators spanning the space of relevant Hamiltonians, $\mathcal{B}_{nj}^{(\mathrm{R})}$ is a basis set of superoperators spanning the space of relevant relaxation superoperators, and $\varphi_{nkj}^{(\mathrm{H,R})}$ are their "phantoms" – arrays of coefficients, one for each Voronoi cell of the mesh, indicating how strong the corresponding interaction or relaxation mechanism is in that particular cell. This is a straightforward extension of the notion of phantom from magnetic resonance imaging.

The superoperators in Eq (28) are generated by *Spinach* on user request; their phantoms are provided by the user. For example, the following *Spinach* syntax is used (3D echo planar imaging example file, Figure 5) to specify longitudinal and transverse relaxation phantoms in an MRI simulation:

```
% Phantom library call and sample settings
[R1_Ph,R2_Ph,PD_Ph,dims,npts]=phantoms('brain-highres');
```

```
                parameters.dims=dims; parameters.npts=npts;

                % Relaxation phantom
                [R1,R2]=rlx_t1_t2(spin_system);
                parameters.rlx_op={R1,R2};
                parameters.rlx_ph={R1_Ph,R2_Ph};

                % Initial and detection state phantoms
                parameters.rho0_ph={PD_Ph};
                parameters.rho0_st={state(spin_system,'Lz','1H')};
                parameters.coil_ph={ones(prod(parameters.npts,1))};
                parameters.coil_st={state(spin_system,'L+','1H')};
```

In the first paragraph, three-dimensional phantoms (longitudinal relaxation rate, transverse relaxation rate, proton density) are requested from the phantom library. The second paragraph requests relaxation superoperators *Spinach* and matches them (`rlx_op`) to their phantoms (`rlx_ph`). In the second paragraph, the initial state (`rho0_st`) is set to longitudinal magnetisation weighted by the proton density in three dimensions (`rho0_ph`). The detection state is set to $\mathbf{L}_+$ uniformly across the sample.

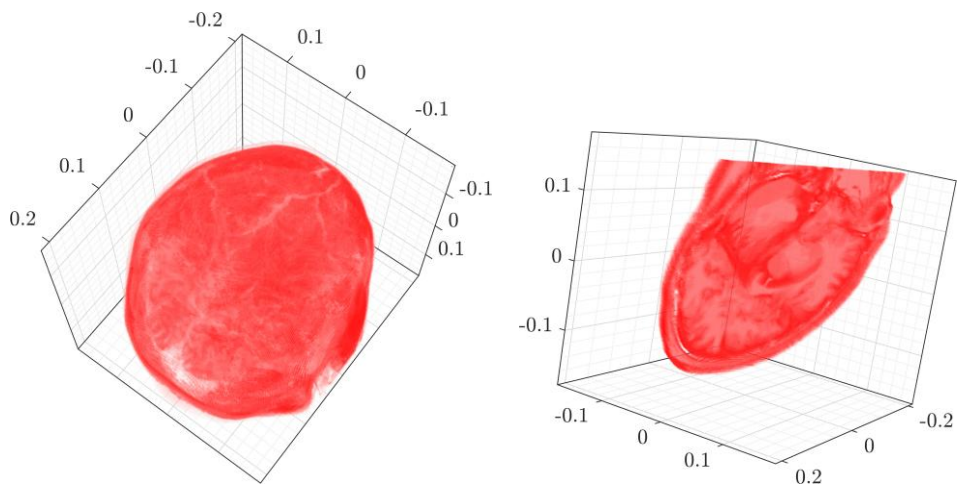

*Figure 5.* A three-dimensional relaxation rate phantom used in medical imaging pulse sequence simulations (from Spinach MRI example set [73], axis labels are in metres). **Left panel:** volumetric plot of the three-dimensional (216 × 180 × 180 voxel grid) distribution of transverse relaxation rates of water protons in a human brain. Water in blood vessels and cerebrospinal fluid pools appears white because it has slower transverse relaxation. **Right panel:** magnetisation excitation slice (after a Gaussian radiofrequency pulse and an echo stage under a magnetic field gradient) showing proton density distribution within the slice. Bone tissue appears white because it has lower proton concentration.

Simulations involving diffusion and flow additionally have a kinetics phantom: an array of coefficients in front of chemical kinetics generators in every Voronoi cell of the sample. This phantom is updated during the simulation because (as discussed in Section 2.3 above) chemical transport rates for nuclear spin states are concentration-dependent. Velocity maps and diffusion tensor maps are provided in a similar way: as arrays of vectors or tensors in every voxel or Voronoi cell.

A unique feature of *Matlab* when applied to physical sciences is that annotated code is often shorter and easier to understand than its verbal description; the code implementing the content of this section is available on GitHub under the MIT license in versions 2.10 and later of *Spinach* package.

## 3.5 Solver, concentration stage

The first stage of the simulation involves spatial and chemical degrees of freedom. Their dynamics does not depend on the nuclear spin state and may therefore be precomputed by solving the corresponding system of differential equations in the time domain:

$$\frac{\partial c_n^{(k)}}{\partial t} = \sum_m F_{km} c_n^{(m)} + f_n\left(c_1^{(k)}, \ldots, c_N^{(k)}\right) \tag{29}$$

where the $c_n^{(k)}$ is the concentration of substance $n$ in Voronoi cell $k$, $\mathbf{F}$ is the transport matrix obtained as described in Section 3.2, and $f_n\left(c_1^{(k)}, \ldots, c_N^{(k)}\right)$ is the right hand side of the mass action law in Eq (1) describing the kinetics of substance $n$.

For elementary liquid-phase chemical reactions, the right hand side of Eqs (29) is always either a linear, or bilinear, or quadratic polynomial function of concentrations. This system is therefore well-behaved and may be solved using standard methods, for example Runge-Kutta [104,105]. For aesthetic reasons (chemical kinetics is a Lie semigroup action), we use Lie group solvers [92,93,106]. The result is the time dependence $c_n^{(k)}(t)$ of the concentration of each substance in each Voronoi cell of the mesh. These concentrations determine the rate multipliers in front of the chemical transport generators in Eq (9) and allow us to proceed to the spin dynamics part.

## 3.6 Solver, nuclear spin stage

Once the concentration dynamics is known from solving Eq (29) and assumed to be nuclear state independent, Eq (22) is reduced to a form in which the evolution generator depends on time and location but not on the concentration-weighted density matrix. Our remaining tasks are to assemble the combined generator to match the state vector structure discussed in Section 3.4, to run the time evolution, and to project out the observable quantities at each spatial location.

The principal challenge here is astronomical matrix dimensions: for the reactions discussed in the examples below (27 proton spins), even the reduced Liouville space has dimension exceeding 30,000. When this is combined with diffusion and flow across tens of thousands of Voronoi cells (the modest case of the microfluidic chip shown in the right panel of Figure 1 has 28,902 cells), the combined dimension of the problem goes into billions and becomes intractable even with sparse matrix arithmetic on the strongest existing GPUs. Thankfully, a workaround was recently published [73] that uses relations of the following type [107]:

$$[\mathbf{A} \otimes \mathbf{B}] \mathbf{v} = \text{vec}\left[\mathbf{B} \mathbf{V} \mathbf{A}^{\text{T}}\right] \tag{30}$$

where $\mathbf{A}$, $\mathbf{B}$, $\mathbf{V}$ are matrices, $\mathbf{v}$ is a vector, and $\text{vec}$ stands for *vectorisation* – a column-wise reshape of a matrix into a vector. The matrix $\mathbf{V}$ is obtained by the reverse procedure: cutting up the column vector $\mathbf{v}$ into strips of appropriate size and concatenating them in the horizontal dimension. For large matrices $\mathbf{A}$ and $\mathbf{B}$, the right hand side of Eq (30) requires significantly less memory because the Kronecker product $\mathbf{A} \otimes \mathbf{B}$ is never computed explicitly. This method may be extended to linear combinations of Kronecker products of any number of matrices [73]:

$$\left(\alpha[\mathbf{A} \otimes \mathbf{B} \otimes \ldots] + \beta[\mathbf{C} \otimes \mathbf{D} \otimes \ldots] + \ldots\right)\mathbf{v} = \alpha[\mathbf{A} \otimes \mathbf{B} \otimes \ldots]\mathbf{v} + \beta[\mathbf{C} \otimes \mathbf{D} \otimes \ldots]\mathbf{v} + \ldots \tag{31}$$

and thus to any evolution generator within the remit of this work. *Spinach* includes a dedicated object that pretends (to *Matlab*) to be a matrix, but instead buffers linear combinations of unevaluated Kronecker products for the purposes of running Eq (31) every time its action on a vector is needed. With this technicality out of the way, we proceed in the following stages at each step of a time-domain simulation of the combined dynamics:

1. Use pre-computed concentrations of each substance in each Voronoi cell of the mesh to calculate chemical kinetics superoperators in each cell using Eq (6). Concatenate the superoperators into a block-diagonal sparse matrix matching the state vector structure in Eq (24).

2. Use interaction and relaxation phantom information to assemble spin Hamiltonian commutation superoperators and relaxation superoperators in each cell. Concatenate the superoperators into a block-diagonal sparse matrix matching the state vector structure in Eq (24). Elementary spin operators are not location- or time-dependent (only their coefficients are) and may therefore be precomputed. If the Hamiltonian or the relaxation superoperator are time- or location-independent, they may also be pre-computed.

3. Assemble the full evolution generator by adding up the kinetics superoperator from Item 1, spin evolution generators from Item 2, and the transport generator $\mathbf{F} \otimes \mathbf{1}$ that influences location coordinates but has no effect on the spin state.

4. Call the time propagation function (`step.m` in *Spinach*) that calculates the exponential action of the evolution generator on the state vector without explicitly exponentiating the generator using one of the many variations of the Krylov method (see Section 4.9.6 in [108]).

5. Use Eq (27) with pre-computed (using user-supplied $b_+$ maps) detection state vectors to calculate appropriate observables in each cell of the mesh.

Polyadic objects [73,109] should be used whenever possible to avoid storing, manipulating, and acting by identical copies of operators. In all but the smallest cases, the use of FP64-capable GPUs is essential because array dimensions go into many millions; *Spinach* does that automatically on user request.

## 4. Examples and benchmarks

In this section, we present practical applications of the formalism built above in the order of increasing feature complexity: from individual processes (diffusion, flow, kinetics, spin evolution), to their combinations, and then to composite simulations involving all types of dynamics simultaneously.

### 4.1 Non-linear kinetics

A convenient class of second-order reactions that can run in seconds to minutes, and therefore may be followed in real time by NMR, is Diels-Alder cycloaddition [110]. Here we consider the reaction between 1,3-cyclopentadiene (A) and acrylonitrile (B) that yields two enantiomeric pairs of bicyclo[2.2.1]hept-5-ene-2-carbonitrile isomers [111]. We call them endo- and exo-norbornene carbonitrile (NBCN) and denote them (C) and (D), respectively:

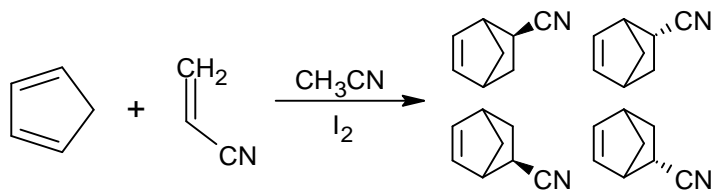

Although chemically identical in achiral environments, the enantiomers may still have to be declared as distinct products in the reaction specification (Section 3.2) because the initial cyclopentadiene may in principle have a non-symmetric nuclear spin state. Here we use symmetric initial spin states and therefore the reaction products are endo- and exo-norbornene carbonitrile:

$$[A] + [B] \xrightarrow{k_1} [C]$$
$$[A] + [B] \xrightarrow{k_2} [D] \quad (32)$$

The rate equations may be written in a form that is not commonly used in chemistry textbooks, but presents an instance of the Lie equation [112] with a state-dependent evolution generator that fits neatly into the algebraic form imposed by Eq (22):

$$\frac{d}{dt}\begin{pmatrix}[A]\\[B]\\[C]\\[D]\end{pmatrix} = \begin{pmatrix} -(k_1+k_2)[B] & 0 & 0 & 0 \\ 0 & -(k_1+k_2)[A] & 0 & 0 \\ 0 & k_1[A] & 0 & 0 \\ 0 & k_2[A] & 0 & 0 \end{pmatrix}\begin{pmatrix}[A]\\[B]\\[C]\\[D]\end{pmatrix} \quad (33)$$

An example solution, obtained using the state-dependent geometric integrator introduced in [95] and already available in *Spinach* [108], is given in the left panel of Figure 6.

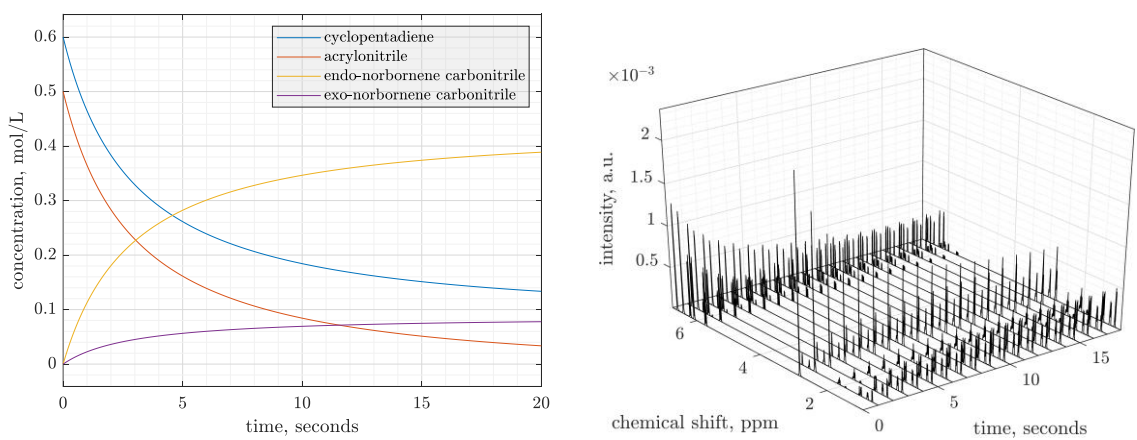

*Figure 6. (Left panel)* An example solution of the rate equations describing the reactions in Eq (32) and Figure 5 with $k_1$ = 250 mol/(L·s), $k_2$ = 50 mol/(L·s), $A_0$ = 0.6 mol/L, $B_0$ = 0.5 mol/L, $C_0 = D_0 = 0$. The solution was obtained using a state-dependent geometric integrator [92,106] because it is compatible with subsequent spin evolution calculations, but any standard method (e.g. Runge-Kutta [104,105]) may also be used at this stage. *(Right panel)* Simulated kinetic profile corresponding to a small flip angle pulse-acquire NMR experiment being performed on the reaction mixture every second for 18 seconds. The simulation was done by time-domain propagation, including Bloch-Redfield-Wangsness relaxation theory, of the concentration-weighted density matrix in Liouville space as described in the main text. Distances (for dipolar interaction tensors) and chemical shielding tensors required by the relaxation superoperator were estimated using density functional theory (GIAO M06/cc-pVTZ in SMD chloroform using Gaussian16), rotational correlation times were estimated using Stokes-Einstein equation.

At this point, we have textbook chemical kinetics [113] that serves as a unit test on the way to the more complicated composite dynamics cases below.

## 4.2 Diffusion and flow

The other textbook limit is pure spatial transport without kinetics or quantum dynamics, corresponding to Eq (21) running forward in the time domain through a particular mesh of Voronoi cells. This is standard [98,99] and therefore used as a unit test in this work; Figure 7 shows two examples: flow through the sample chamber of a microfluidic chip with a no-slip boundary condition (left panel) and diffusion from a localised initial condition on the same grid. The velocity profile was imported from COMSOL; the mesh was pre-processed as described in Section 3.1 above.

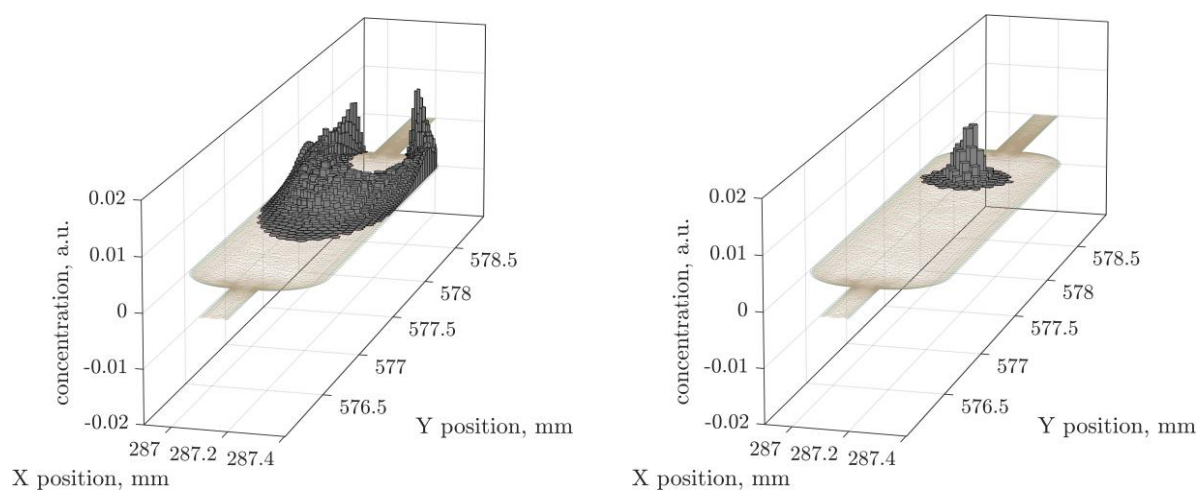

*Figure 7. Stop-frames of flow and diffusion simulations in the sample chamber of the microfluidic chip in Figure 1 (simulation and visualisation scripts are included with the example set of Spinach 2.10 and later). Substance concentrations in each Voronoi cell are shown as heights of the grey columns. **Left:** flow under the stationary velocity field computed by COMSOL with the initial concentration set to a non-zero value in the distal pipe. **Right:** diffusion from a non-zero concentration in a single Voronoi cell in the middle of the chip. Full videos are in the Supplementary Information.*

## 4.3 Non-linear kinetics with spin evolution

We now come to the boundary of the published prior art [23] – a combination of non-linear kinetics and spin dynamics that goes beyond what may be described by Bloch equations, the improvement being that our formalism is numerically stable. At this point, the procedures described in Section 2.3 and 3.3 must be performed for the particular cycloaddition reaction discussed in Section 4.1.

The necessary spin indexing is illustrated in Figure 8. In the *Spinach* input script, all reactants and products are specified in the same input stream and then partitioned into sets of spins belonging to individual substances as described in Section 3.3: spins 1-6 for cyclopentadiene, 7-9 for acrylonitrile, and so on. The matching table shown in Figure 8 is then supplied, it tells the reaction generator function which spin on the left side of the reaction arrow becomes which spin on the right in each chemical process. Here we have two reactions – one producing exo- and the other endo-isomer of norbornene carbonitrile – and therefore two matching tables.

The spin evolution generator is composed of the Hamiltonian commutation superoperator (we have used experimentally determined chemical shifts and *J*-couplings) and the relaxation superoperator that was obtained using Bloch-Redfield-Wangsness theory. Its numerical implementation technicalities are reported elsewhere [114,115]; here we have included dipolar relaxation, chemical shift anisotropy (CSA) relaxation, and their cross-correlations. Molecular geometries and anisotropic parts of the chemical shielding tensors were obtained from DFT calculations (GIAO M06/cc-pVTZ in SMD acetonitrile) using *Gaussian* [116-120] and imported into *Spinach*. Proton shielding tensors for one of the products are shown in Figure 9.

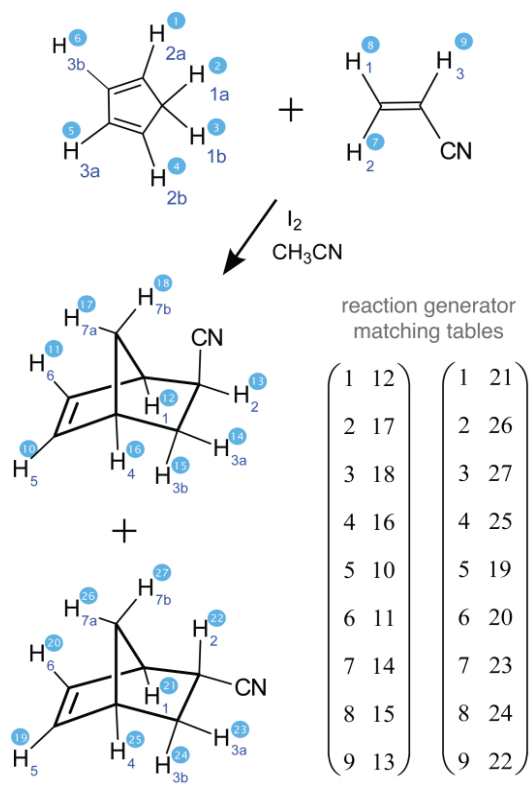

*Figure 8. Diels-Alder cycloaddition reaction schematic with an illustration of reaction generator matching table buildling algorithm. The process is treated as two reactions, one leading to endo-isomer, the other to exo-isomer of norbornene carbonitrile. Accordingly, two matching tables are needed for which spin on the left hand side of the reaction arrow becomes which spin on the right hand side.*

The use of quantum state transport generators is best illustrated using *Matlab* code. We first send reaction specifications described in Section 3.3 to the generator build script:

```
% Reaction generators
G1=react_gen(spin_system,reaction{1});
G2=react_gen(spin_system,reaction{2});
```

here, `spin_system` is the global object used by *Spinach* to store spin system information [103]. The resulting generator variable is a cell array of matrices, one per reactant, draining and mapping each basis state in the reactant state space into its destination in the product state space.

The kinetic terms of Eq (22) may now be built. The state vector is concentration-weighted – for each reactant, its own concentration is already in the state vector – and therefore only concentrations of the other species should be present in the coefficients multiplying the generators:

```
% Build the composite evolution generator
F=H+1i*R+1i*k1*G1{1}*B(t) ...    % Reaction 1 from substance A
   +1i*k1*A(t)*G1{2} ...         % Reaction 1 from substance B
   +1i*k2*G2{1}*B(t) ...         % Reaction 2 from substance A
   +1i*k2*A(t)*G2{2};            % Reaction 2 from substance B
```

In this expression, concentrations come from the preceding calculation (Section 3.5) of kinetics and spatial transport. Note the absence of concentration denominators everywhere – this formalism is numerically stable at low concentrations. At this point, the evolution generator assembly is finished and its exponential action may be used to propagate the system forward in time.

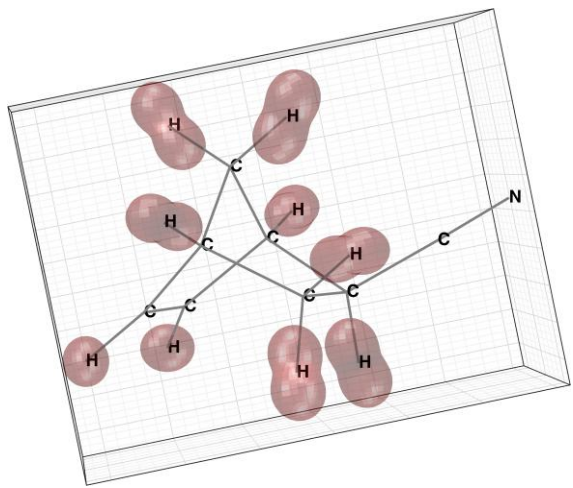

*Figure 9. Visual representation of the absolute proton chemical shielding tensors (CSTs) in exo-norbornene carbonitrile. CSTs were obtained as described in the main text and used for the calculation of relaxation superoperators. They are plotted here using the spherical harmonic visualiser described in Section 3.3.4 of [15] and supplied with Spinach.*

The result is shown in the right panel of Figure 6; this is the simplest calculation that requires the use of the two-stage process described in Section 3 – first the concentrations, then the spin dynamics using those concentrations as known functions of time. The first stage was discussed above, the structure of the second stage time loop is best illustrated with *Matlab* code:

```
% Preallocate the trajectory and get it started
traj=zeros([numel(eta) nsteps+1]); traj(:,1)=eta;

% Run evolution
for n=1:nsteps

    % Build the left interval edge composite evolution generator
    F_L=1i*k1*G1{1}*B(time_axis(n))...    % Reaction 1 from substance A
       +1i*k1*A(time_axis(n))*G1{2}...    % Reaction 1 from substance B
       +1i*k2*G2{1}*B(time_axis(n))...    % Reaction 2 from substance A
       +1i*k2*A(time_axis(n))*G2{2};      % Reaction 2 from substance B

    % Build the right interval edge composite evolution generator
    F_R=1i*k1*G1{1}*B(time_axis(n+1))... % Reaction 1 from substance A
       +1i*k1*A(time_axis(n+1))*G1{2}... % Reaction 1 from substance B
       +1i*k2*G2{1}*B(time_axis(n+1))... % Reaction 2 from substance A
       +1i*k2*A(time_axis(n+1))*G2{2};   % Reaction 2 from substance B

    % Take the time step using the two-point Lie quadrature
    traj(:,n+1)=step(spin_system,{F_L,F_R},traj(:,n),dt);

end
```

Here, the time is sampled from a finite grid on which the spin evolution trajectory is then calculated using the two-point product quadrature [92,106] that was recently implemented into *Spinach* [95].

## 4.4 Diffusion and flow with spin evolution

This special case is well researched and covered in the literature – diffusion MRI for Bloch equation models [9,121], diffusion-ordered NMR spectroscopy for large coupled spin systems [122,123], rotational [124,125] and translational [126] diffusion as relaxation mechanisms, material porosity characterisation by long-lived state diffusion measurements [127], *etc.* The mathematics here is a straightforward direct product of spin and location degrees of freedom [72,125], its implementation in *Spinach* has already been discussed elsewhere [128,129]. The only major numerical simulation problem in these settings – large matrix dimensions – has recently been solved [73]. For our purposes here, this is a unit test on the way to more complicated simulations.

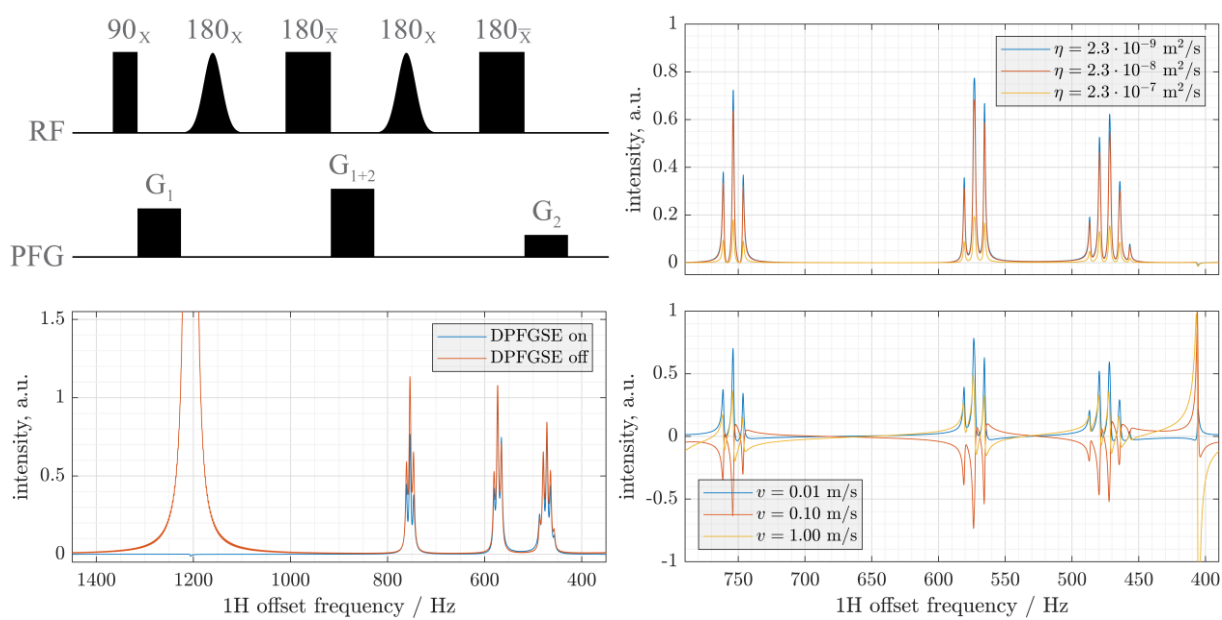

*Figure 10.* Double pulsed field gradient spin echo (DPFGSE) pulse sequence and its simulated performance for a coupled six-spin system of gamma-aminobutyric acid (GABA) dissolved in 95% $D_2O$, with the spatial dynamics (diffusion and flow) calculations performed on an explicit spatial grid representing a 15 mm long NMR sample as described in the main text. **Top left:** DPFGSE pulse sequence, selective Gaussian pulses are applied to the water signal, RF = radiofrequency, PFG = pulsed field gradient. **Bottom left:** 250 MHz $^1$H NMR spectrum of GABA with (blue line) and without (red line) DPFGSE water signal suppression; note the attenuation of the signal at 1200 Hz. **Top right:** useful signal attenuation resulting from incomplete magnetisation refocussing, as a function of the diffusion coefficient; blue line corresponds to water at room temperature. **Bottom right:** signal phase errors resulting from the presence of hydrodynamic flow with indicated velocities. The artefact at 400 Hz is a reflection of the incompletely suppressed water signal.

In the context of dynamics of multi-spin systems, a good illustration of the implementation reported here is selective suppression and excitation of NMR signals using the double pulsed field gradient spin echo (DPFGSE [130]) pulse sequence (Figure 10, top left). This is one of the best solvent signal suppression methods because its mechanism is resistant to gradient spiral refocussing errors introduced by diffusion (Figure 10, top right). Here, spatial dynamics is generated by the first derivative operator with respect to location (hydrodynamic flow) and second derivative operator (translational diffusion); both were represented by finite difference operators generated as described in [73] using seven-point stencils on a 500-point grid for a 15 mm long sample with a periodic boundary condition.

The simulations were done for a 5.87 Tesla magnet (250 MHz proton frequency), using 0.10 T/m ($G_1$) and 0.15 T/m ($G_2$) pulsed field gradients of 1.0 ms duration, explicit 20 ms Gaussian soft pulses with 1220 Hz offset, nutation frequency of 1700 Hz, and 10 discretisation slices. Diffusion coefficients and flow velocities were varied as shown in Figure 10. Physically correct outcomes are seen: accelerating diffusion (a spatially symmetric process) causes magnetisation losses but no artefacts (top right panel), whereas accelerating the flow (a non-symmetric process) degrades solvent suppression performance and generates phase distortions due to incomplete refocussing of the gradient spirals.

**4.5 Diffusion and flow with non-linear kinetics and spin evolution**

This is our final destination; all processes described in the previous sections are now simultaneously present. The principal problem here is that the combined dimension of spin evolution generators for the cycloaddition reaction (30,466 in the IK-2 basis set [85]) and spatial dynamics generators for diffusion and flow (2659 cells for the mesh shown in Figure 7) is close to 100 million – this makes the polyadic representation for the combined evolution generator [73] unavoidable. Strong GPUs (here we use a server with eight Nvidia H200 cards) are also recommended.

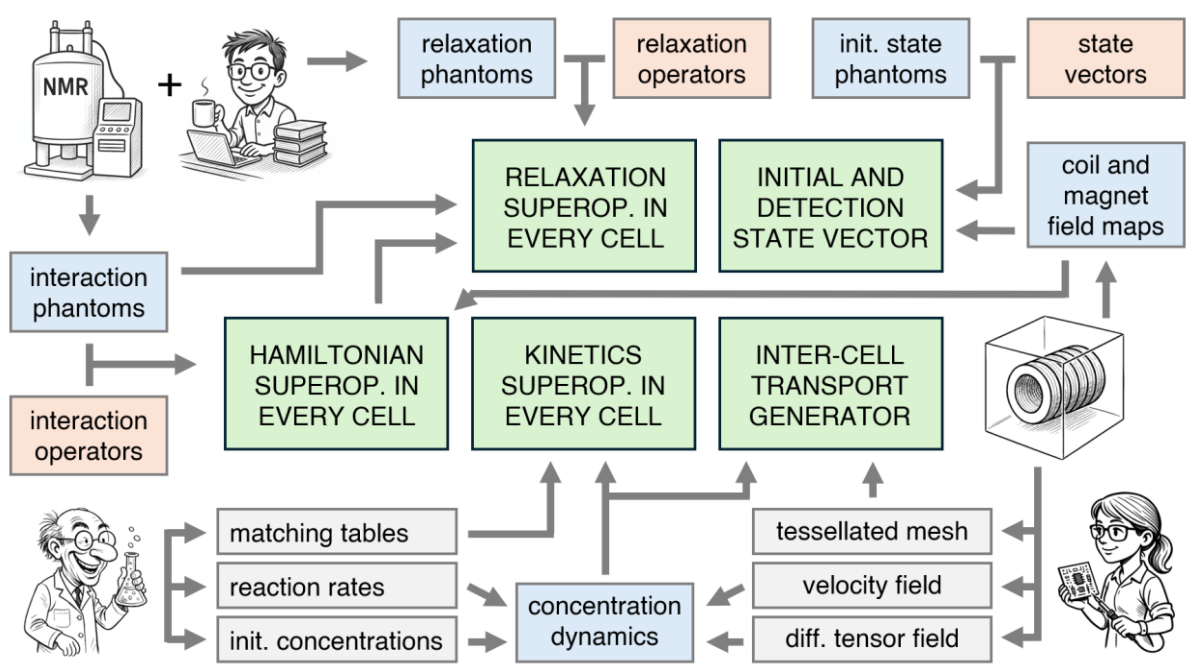

*Figure 11. General overview of the code flow in the microfluidics module of Spinach. There are five types of phantoms (blue squares) that specify spatial distributions of simulation components: interaction (e.g. pulsed field gradients), relaxation (e.g. tissue type), initial state (e.g. initial magnetisation distribution), coil field map (obtained, along with the flow velocity field, from separate COMSOL simulations), and concentration (e.g. chemical reactions and spatial transport). These phantoms are combined with their corresponding Liouville space superoperators as described in the main text to produce the overall system evolution generator at each time step in the simulation. Concentration dynamics is precomputed separately and then used as a set of time-dependent parameters in the spatially distributed spin dynamics simulation. Any similarity between the cartoon characters and any of the authors is absolutely intentional.*

The calculation proceeded in the following stages (an overview schematic is given in Figure 11, the process is automated in *Spinach* 2.11 and later versions):

1. **Concentration dynamics simulation** from the user-specified initial concentration distribution (small amounts of cyclopentadiene and acrylonitrile in adjacent regions of the reaction chamber, Figure 12) in the presence of kinetics and spatial motion:

(a) Diffusion and flow evolution generator $\mathbf{F}$ is built as described in Section 3.2, the polynomials $f_n\left(c_1^{(k)},\ldots,c_N^{(k)}\right)$ responsible for the reaction kinetics are assembled from the reaction descriptors (Section 3.3) provided by the user.

(b) The two-point Lie group method [92-94,106] implemented in *Spinach* `step.m` function [95] is used to solve Eq (29) and obtain the time dependence of all concentrations in all Voronoi cells of the mesh (Figure 12).

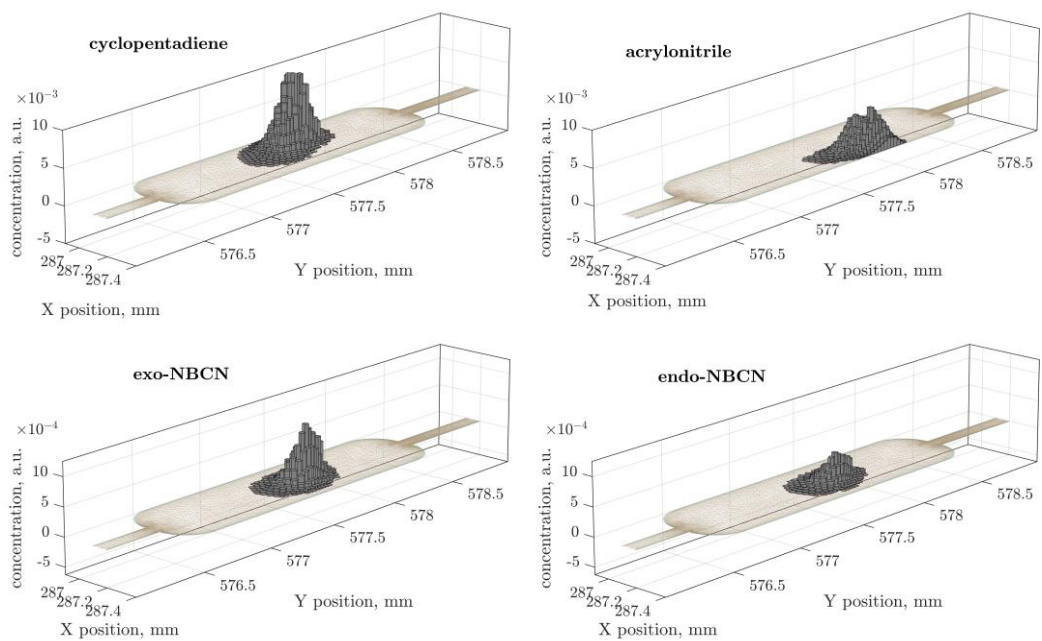

*Figure 12. A frame from the spatial dynamics and chemical kinetics simulation stage. The initial condition is drops of cyclopentadiene and acrylonitrile in adjacent regions in the upper part of the chip. As the reagents flow downwards and mix, the reaction produces unequal quantities of exo- and endo-norbornene carbonitrile. The resulting time dependence of all concentrations in all Voronoi cells of the mesh is used in the subsequent stages to generate the non-linear kinetics superoperator at every time point in the spin dynamics simulation.*

(c) An array of *Matlab* `griddedInterpolant` objects (one for each substance in each cell) is created. Their role is to interpolate time – the time grid used by the subsequent spin dynamics simulation stage is not necessarily the same.

2. **Spin dynamics infrastructure assembly** must account for the fact that some terms in Eq (22) may be time-, location-, and concentration-dependent:

(a) Chemical reaction generators are built as described in Section 2.3 above. At this point, those are generators in the Lie algebraic sense; they will be multiplied by appropriate coefficients when the time evolution loop is computed.

(b) Superoperators of all pertinent individual interactions (including Cartesian spin operators for use in pulses and pulsed field gradients) are requested from *Spinach* kernel; their coefficients will also be decided when the time evolution loop starts.

(c) All pertinent individual spin state vectors are requested from *Spinach* kernel. The initial state is assembled in every cell of the mesh by multiplying these vectors by appropriate concentrations as per Eq (24). Detection states are built using the same state vectors, but using coil field maps as coefficients as per Eq (26): the coil both sees and affects different locations of the mesh with different coefficients [12,16].

3. **Time evolution loop** must, at each step, apply the various time-, location-, and concentration dependent coefficients to the infrastructure operators built above, assemble the overall evolution generator, and propagate the system forward in time. At each time step:

   (a) Diffusion and flow generator matrix $\mathbf{F}$ is inherited from Stage 1. Spatial dynamics is assumed to be nuclear spin independent; a Kronecker product with a unit spin superoperator must therefore be taken. In practice, $\mathbf{F} \otimes \mathbf{1}$ is stored as a polyadic object and the Kronecker product is never computed explicitly. If $\mathbf{F}$ is time-independent, this part may be pre-computed in Stage 2.

   (b) Hamiltonian, relaxation, and kinetics superoperator components are assembled in each cell of the mesh with appropriate time- and location-dependent coefficients describing radiofrequency pulses, pulsed field gradients, concentrations, *etc.* Time-independent terms may be pre-computed in Stage 2. This completes the building of the spin evolution generator in each cell of the mesh.

   (c) Spin evolution generators pertaining to individual cells of the mesh are concatenated into a block-diagonal matrix. Location-independent terms may be stored as $\mathbf{1} \otimes \mathbf{H}$ and $\mathbf{1} \otimes \mathbf{R}$ polyadic terms with un-opened Kronecker products. This completes the construction of the global evolution generator acting on the column of concentration-weighted state vectors in Eq (22).

   (d) At this point, typical matrix dimensions are in the millions (thousands for the mesh cell count kroneckered with thousands for the spin state space) – what saves us from a memory overflow is the fact that both sets of matrices are very sparse and some are stored in a polyadic format. Exponentiating such an object explicitly is out of the question – the time step must be computed using a Krylov-like method, we recommend the one described in Section 4.9.6 of IK's book [108].

   (e) The propagated state vector may either be stored for later use, or observables may be computed at this point using Eq (27) if memory efficiency is a concern – unlike the evolution generators, the trajectory array is not sparse.

At this point, we have our detailed microscopic trajectory with cell-by-cell chemical-, hydro-, and spin dynamics, and also the observables seen by the radiofrequency coil – the signals in Figure 13 first rise as chemicals flow into the region where spins are affected and seen by the coil, changing their relative intensity as the chemical reaction proceeds, and then fade as the reaction products and leftover reagents flow out of the coil area. Throughout the process, the quantum mechanical description of spin dynamics is maintained, as evidenced by the *J*-coupling patterns in all NMR signals.

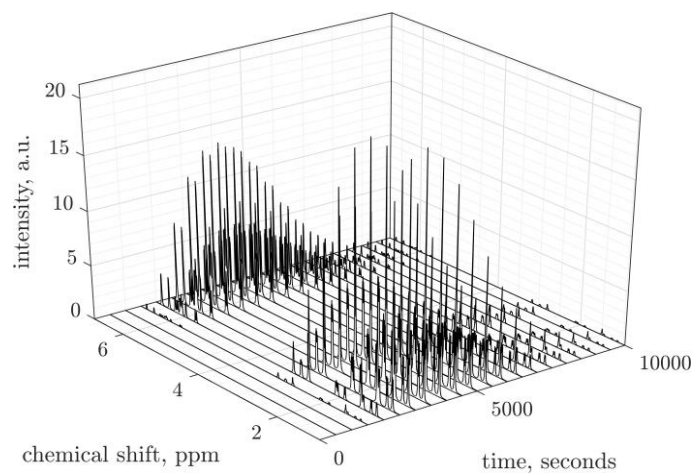

*Figure 13. Simulated pulse-acquire 600 MHz $^1$H NMR spectra of a flowing, diffusing, and reacting Diels-Alder cycloaddition system (acrylonitrile + cyclopentadiene into endo- and exo-norbornene carbonitrile, 30 proton spins in total) as excited and then detected at regular intervals by a radiofrequency coil located between 238.0 and 238.3 mm X position, and between 577.0 and 577.5 mm Y position in the reaction chamber of the microfluidic chip shown in Figures 8 and 13. Reagents are initially located outside the coil, the signals appear as they flow in and change relative intensities as the reaction proceeds. The signals then fade to zero as the products and the remaining reagents flow out of the other end of the coil.*

Although we do not discuss visualisation logistics (the matter is well researched), considerable further programming and data handling effort is of course needed to present time- and location-dependent NMR observables in a user-friendly way. Those infrastructure functions are supplied with *Spinach*.

## Conclusions and outlook

A combination of recent developments in spin dynamics (restricted state spaces [85], polyadic evolution generators [73]), numerical linear algebra (Lie group integrators [92-94,106], spectral methods [131]) and computer science (sparse matrix libraries [132], graphics processing units [133,134], specialised scientific computing languages [96]) has finally made it possible to perform nuclear magnetic resonance simulations for non-trivial spin systems in the simultaneous presence of diffusion, hydrodynamics, and non-linear kinetics. An open-source implementation in *Spinach* has required a formalism update (to avoid a numerical stability issue in the prior art [23]) and considerable software engineering effort; they are described in this paper. As an example, we have used a cycloaddition reaction in a flowing microfluidic chip, for which the simulation is just about feasible on a strong GPU server at the time of writing; it will likely fit into a laptop in a few years' time.

## Acknowledgements

This work was supported by a research grant from the Center for New Scientists at the Weizmann Institute of Science. We are also grateful to the EPSRC (EP/W020343/1) and to the Israeli Army for the missile defence support that has kept IK safe in Rehovot as this paper was being written. AA thanks MathWorks for a studentship grant and for the general awesomeness of their engineering and technical support teams. The authors acknowledge the use of NMRBox [135] and the IRIDIS High

Performance Computing Facility and associated support services at the University of Southampton, as well as the use of Tesla A100 GPUs through NVIDIA Academic Grants Programme.# References

1. Schrödinger, E., *An undulatory theory of the mechanics of atoms and molecules.* Physical Review, 1926. **28**(6): p. 1049.

2. von Neumann, J., *Wahrscheinlichkeitstheoretischer Aufbau der Quantenmechanik.* Nachrichten von der Gesellschaft der Wissenschaften zu Göttingen: Mathematisch-Physikalische Klasse, 1927: p. 245-272.

3. Kuprov, I., *What exactly is spin?*, in *Spin: from basic symmetries to quantum optimal control*. 2023, Springer. p. 43-72.

4. van't Hoff, J.H., *Die grenzebene, ein beitrag zur kenntniss der esterbildung.* Berichte der Deutschen Chemischen Gesellschaft, 1877. **10**(1): p. 669-678.

5. Navier, C., *Mémoire sur les lois du mouvement des fluides*, in *Mémoires de l'Académie Royale des Sciences de l'Institut de France*. 1822, Gauthier-Villars. p. 389-440.

6. Stokes, G.G., *On the theories of the internal friction of fluids in motion and of the equilibrium and motion of elastic solids.* Transactions of the Cambridge Philosophical Society, 1845. **8**: p. 287-319.

7. Kominis, I.K., *Quantum Zeno effect explains magnetic-sensitive radical-ion-pair reactions.* Physical Review E, 2009. **80**(5): p. 056115.

8. Nelson, S.J., J. Kurhanewicz, D.B. Vigneron, P.E.Z. Larson, A.L. Harzstark, M. Ferrone, M. van Criekinge, J.W. Chang, R. Bok, I. Park, G. Reed, L. Carvajal, E.J. Small, P. Munster, V.K. Weinberg, J.H. Ardenkjaer-Larsen, A.P. Chen, R.E. Hurd, L.-I. Odegardstuen, F.J. Robb, J. Tropp, and J.A. Murray, *Metabolic Imaging of Patients with Prostate Cancer Using Hyperpolarized [1-$^{13}$C]Pyruvate.* Science Translational Medicine, 2013. **5**(198): p. 198ra108-198ra108.

9. Topgaard, D., *Multidimensional diffusion MRI.* Journal of magnetic resonance, 2017. **275**: p. 98-113.

10. Bart, J., A.J. Kolkman, A.J. Oosthoek-de Vries, K. Koch, P.J. Nieuwland, H. Janssen, J. van Bentum, K.A. Ampt, F.P. Rutjes, and S.S. Wijmenga, *A microfluidic high-resolution NMR flow probe.* Journal of the American Chemical Society, 2009. **131**(14): p. 5014-5015.

11. Davoodi, H., N. Nordin, L. Bordonali, J.G. Korvink, N. MacKinnon, and V. Badilita, *An NMR-compatible microfluidic platform enabling in situ electrochemistry.* Lab on a Chip, 2020. **20**(17): p. 3202-3212.

12. Finch, G., A. Yilmaz, and M. Utz, *An optimised detector for in-situ high-resolution NMR in microfluidic devices.* Journal of Magnetic Resonance, 2016. **262**: p. 73-80.

13. Born, M., *Zur Quantenmechanik der Stoßvorgänge.* Zeitschrift für Physik, 1926. **37**(12): p. 863-867.

14. Kaptein, R., *Simple rules for chemically induced dynamic nuclear polarization.* Journal of the Chemical Society D: Chemical Communications, 1971(14): p. 732-733.

15. Kuprov, I., *Bestiary of Spin Hamiltonians*, in *Spin: From Basic Symmetries to Quantum Optimal Control*. 2023, Springer. p. 73-105.
27

16. Barker, S.J., L. Dagys, W. Hale, B. Ripka, J. Eills, M. Sharma, M.H. Levitt, and M. Utz, *Direct Production of a Hyperpolarized Metabolite on a Microfluidic Chip.* Analytical Chemistry, 2022. **94**(7): p. 3260-3267.

17. *COMSOL Multiphysics*. COMSOL AB, Stockholm, Sweden.

18. Alexander, S., *Exchange of interacting nuclear spins in nuclear magnetic resonance. II. Chemical exchange.* The Journal of Chemical Physics, 1962. **37**(5): p. 974-980.

19. Bain, A.D., *Chemical exchange in NMR.* Progress in nuclear magnetic resonance spectroscopy, 2003. **43**(3-4): p. 63-103.

20. Brindle, K.M., *NMR methods for measuring enzyme kinetics in vivo.* Progress in Nuclear Magnetic Resonance Spectroscopy, 1988. **20**(3): p. 257-293.

21. Brindle, K.M. and I.D. Campbell, *NMR studies of kinetics in cells and tissues.* Quarterly reviews of biophysics, 1987. **19**(3-4): p. 159-182.

22. Gutowsky, H., R. Vold, and E. Wells, *Theory of chemical exchange effects in magnetic resonance.* The Journal of Chemical Physics, 1965. **43**(11): p. 4107-4125.

23. Kühne, R.O., T. Schaffhauser, A. Wokaun, and R.R. Ernst, *Study of transient chemical reactions by NMR: fast stopped-flow Fourier transform experiments.* Journal of Magnetic Resonance, 1979. **35**(1): p. 39-67.

24. Perrin, C.L. and T.J. Dwyer, *Application of two-dimensional NMR to kinetics of chemical exchange.* Chemical reviews, 1990. **90**(6): p. 935-967.

25. Warburg, O., *On the origin of cancer cells.* Science, 1956. **123**(3191): p. 309-314.

26. Romero-Garcia, S., J.S. Lopez-Gonzalez, J.L. B´ez-Viveros, D. Aguilar-Cazares, and H. Prado-Garcia, *Tumor cell metabolism: an integral view.* Cancer biology & therapy, 2011. **12**(11): p. 939-948.

27. Hesketh, R.L. and K.M. Brindle, *Magnetic resonance imaging of cancer metabolism with hyperpolarized 13C-labeled cell metabolites.* Current opinion in chemical biology, 2018. **45**: p. 187-194.

28. Josan, S., K. Billingsley, J. Orduna, J.M. Park, R. Luong, L. Yu, R. Hurd, A. Pfefferbaum, D. Spielman, and D. Mayer, *Assessing inflammatory liver injury in an acute CCl4 model using dynamic 3D metabolic imaging of hyperpolarized [1-13C] pyruvate.* NMR in Biomedicine, 2015. **28**(12): p. 1671-1677.

29. Nielsen, P.M., C. Laustsen, L.B. Bertelsen, H. Qi, E. Mikkelsen, M.L.V. Kristensen, R. Nørregaard, and H. Stødkilde-Jørgensen, *In situ lactate dehydrogenase activity: a novel renal cortical imaging biomarker of tubular injury?* American Journal of Physiology-Renal Physiology, 2017. **312**(3): p. F465-F473.

30. Nielsen, P.M., H. Qi, L.B. Bertelsen, and C. Laustsen, *Metabolic reprogramming associated with progression of renal ischemia reperfusion injury assessed with hyperpolarized [1-13C] pyruvate.* Scientific reports, 2020. **10**(1): p. 8915.

31. Laustsen, C., J.A. Østergaard, M.H. Lauritzen, R. Nørregaard, S. Bowen, L.V. Søgaard, A. Flyvbjerg, M. Pedersen, and J.H. Ardenkjær-Larsen, *Assessment of early diabetic renal changes with hyperpolarized [1-13C] pyruvate.* Diabetes/metabolism research and reviews, 2013. **29**(2): p. 125-129.

32. Bonavita, S., F. Di Salle, and G. Tedeschi, *Proton MRS in neurological disorders.* European journal of radiology, 1999. **30**(2): p. 125-131.

33. de Graaf, R.A., G.F. Mason, A.B. Patel, K.L. Behar, and D.L. Rothman, *In vivo 1H-[13C]-NMR spectroscopy of cerebral metabolism.* NMR in Biomedicine: An International Journal Devoted to the Development and Application of Magnetic Resonance In Vivo, 2003. **16**(6-7): p. 339-357.

34. Wang, Z.J., M.A. Ohliger, P.E. Larson, J.W. Gordon, R.A. Bok, J. Slater, J.E. Villanueva-Meyer, C.P. Hess, J. Kurhanewicz, and D.B. Vigneron, *Hyperpolarized 13C MRI: state of the art and future directions.* Radiology, 2019. **291**(2): p. 273-284.

35. Stejskal, E., *Use of spin echoes in a pulsed magnetic-field gradient to study anisotropic, restricted diffusion and flow.* The Journal of Chemical Physics, 1965. **43**(10): p. 3597-3603.

36. Stejskal, E.O. and J.E. Tanner, *Spin diffusion measurements: spin echoes in the presence of a time-dependent field gradient.* The journal of chemical physics, 1965. **42**(1): p. 288-292.

37. Wesbey, G.E., M.E. Moseley, and R.L. Ehman, *Translational molecular self-diffusion in magnetic resonance imaging. I. Effects on observed spin-spin relaxation.* Investigative radiology, 1984. **19**(6): p. 484-490.

38. Basser, P.J., J. Mattiello, and D. LeBihan, *MR diffusion tensor spectroscopy and imaging.* Biophysical journal, 1994. **66**(1): p. 259-267.

39. Bryant, D., J. Payne, D.N. Firmin, and D.B. Longmore, *Measurement of flow with NMR imaging using a gradient pulse and phase difference technique.* J Comput Assist Tomogr, 1984. **8**(4): p. 588-593.

40. Firmin, D., G. Nayler, R. Klipstein, S. Underwood, R. Rees, and D. Longmore, *In vivo validation of MR velocity imaging.* J Comput Assist Tomogr, 1987. **11**(5): p. 751-756.

41. Gatehouse, P.D., J. Keegan, L.A. Crowe, S. Masood, R.H. Mohiaddin, K.-F. Kreitner, and D.N. Firmin, *Applications of phase-contrast flow and velocity imaging in cardiovascular MRI.* European radiology, 2005. **15**: p. 2172-2184.

42. van der Toorn, A., R.M. Dijkhuizen, C.A. Tulleken, and K. Nicolay, *Diffusion of metabolites in normal and ischemic rat brain measured by localized 1H MRS.* Magnetic resonance in medicine, 1996. **36**(6): p. 914-922.

43. Futterer, J.J., S.W. Heijmink, T.W. Scheenen, J. Veltman, H.J. Huisman, P. Vos, C.A.H.V. de Kaa, J.A. Witjes, P.F. Krabbe, and A. Heerschap, *Prostate cancer localization with dynamic contrast-enhanced MR imaging and proton MR spectroscopic imaging.* Radiology, 2006. **241**(2): p. 449-458.

44. Wilson, M., O. Andronesi, P.B. Barker, R. Bartha, A. Bizzi, P.J. Bolan, K.M. Brindle, I.Y. Choi, C. Cudalbu, and U. Dydak, *Methodological consensus on clinical proton MRS of the brain: review and recommendations.* Magnetic resonance in medicine, 2019. **82**(2): p. 527-550.

45. Emwas, A.-H., R. Roy, R.T. McKay, L. Tenori, E. Saccenti, G.N. Gowda, D. Raftery, F. Alahmari, L. Jaremko, and M. Jaremko, *NMR spectroscopy for metabolomics research.* Metabolites, 2019. **9**(7): p. 123.

46. Golman, K., R.i.t. Zandt, M. Lerche, R. Pehrson, and J.H. Ardenkjaer-Larsen, *Metabolic imaging by hyperpolarized 13C magnetic resonance imaging for in vivo tumor diagnosis.* Cancer research, 2006. **66**(22): p. 10855-10860.

47. Brindle, K.M., *Imaging metabolism with hyperpolarized 13C-labeled cell substrates.* Journal of the American Chemical Society, 2015. **137**(20): p. 6418-6427.

48. Bowers, C.R. and D.P. Weitekamp, *Transformation of symmetrization order to nuclear-spin magnetization by chemical reaction and nuclear magnetic resonance.* Physical Review Letters, 1986. **57**(21): p. 2645.

49. Bowers, C.R. and D.P. Weitekamp, *Parahydrogen and synthesis allow dramatically enhanced nuclear alignment.* Journal of the American Chemical Society, 1987. **109**(18): p. 5541-5542.

50. Overhauser, A.W., *Polarization of nuclei in metals.* Physical Review, 1953. **92**(2): p. 411.

51. Carver, T.R. and C.P. Slichter, *Polarization of nuclear spins in metals.* Physical Review, 1953. **92**(1): p. 212.

52. Grist, J.T., M.A. McLean, F. Riemer, R.F. Schulte, S.S. Deen, F. Zaccagna, R. Woitek, C.J. Daniels, J.D. Kaggie, and T. Matys, *Quantifying normal human brain metabolism using hyperpolarized [1–13C] pyruvate and magnetic resonance imaging.* Neuroimage, 2019. **189**: p. 171-179.

53. Kurhanewicz, J., D.B. Vigneron, J.H. Ardenkjaer-Larsen, J.A. Bankson, K. Brindle, C.H. Cunningham, F.A. Gallagher, K.R. Keshari, A. Kjaer, and C. Laustsen, *Hyperpolarized 13C MRI: path to clinical translation in oncology.* Neoplasia, 2019. **21**(1): p. 1-16.

54. Marco-Rius, I., M.C. Tayler, M.I. Kettunen, T.J. Larkin, K.N. Timm, E.M. Serrao, T.B. Rodrigues, G. Pileio, J.H. Ardenkjaer-Larsen, and M.H. Levitt, *Hyperpolarized singlet lifetimes of pyruvate in human blood and in the mouse.* NMR in Biomedicine, 2013. **26**(12): p. 1696-1704.

55. Bak, M., J.T. Rasmussen, and N.C. Nielsen, *SIMPSON: a general simulation program for solid-state NMR spectroscopy.* Journal of magnetic resonance, 2011. **213**(2): p. 366-400.

56. Smith, S., T. Levante, B.H. Meier, and R.R. Ernst, *Computer simulations in magnetic resonance: an object-oriented programming approach.* Journal of Magnetic Resonance, Series A, 1994. **106**(1): p. 75-105.

57. Veshtort, M. and R.G. Griffin, *SPINEVOLUTION: a powerful tool for the simulation of solid and liquid state NMR experiments.* Journal of Magnetic Resonance, 2006. **178**(2): p. 248-282.

58. Benoit-Cattin, H., G. Collewet, B. Belaroussi, H. Saint-Jalmes, and C. Odet, *The SIMRI project: a versatile and interactive MRI simulator.* Journal of Magnetic Resonance, 2005. **173**(1): p. 97-115.

59. Xanthis, C.G. and A.H. Aletras, *coreMRI: A high-performance, publicly available MR simulation platform on the cloud.* PLoS One, 2019. **14**(5): p. e0216594.

60. Stöcker, T., K. Vahedipour, D. Pflugfelder, and N.J. Shah, *High-performance computing MRI simulations.* Magnetic resonance in medicine, 2010. **64**(1): p. 186-193.

61. Xanthis, C.G., I.E. Venetis, A. Chalkias, and A.H. Aletras, *MRISIMUL: a GPU-based parallel approach to MRI simulations.* IEEE Transactions on Medical Imaging, 2013. **33**(3): p. 607-617.

62. Torrey, H.C., *Bloch equations with diffusion terms.* Physical review, 1956. **104**(3): p. 563.

63. Liu, F., J.V. Velikina, W.F. Block, R. Kijowski, and A.A. Samsonov, *Fast realistic MRI simulations based on generalized multi-pool exchange tissue model.* IEEE transactions on medical imaging, 2016. **36**(2): p. 527-537.

64. McConnell, H.M., *Reaction rates by nuclear magnetic resonance.* The Journal of chemical physics, 1958. **28**(3): p. 430-431.

65. Singh, A., A. Debnath, K. Cai, P. Bagga, M. Haris, H. Hariharan, and R. Reddy, *Evaluating the feasibility of creatine-weighted CEST MRI in human brain at 7 T using a Z-spectral fitting approach.* NMR in Biomedicine, 2019. **32**(12): p. e4176.

66. Schuenke, P., D. Paech, C. Koehler, J. Windschuh, P. Bachert, M.E. Ladd, H.-P. Schlemmer, A. Radbruch, and M. Zaiss, *Fast and quantitative T1ρ-weighted dynamic glucose enhanced MRI.* Scientific reports, 2017. **7**(1): p. 42093.

67. Sun, P.Z., *Quasi-steady-state chemical exchange saturation transfer (QUASS CEST) MRI analysis enables T1 normalized CEST quantification–Insight into T1 contribution to CEST measurement.* Journal of Magnetic Resonance, 2021. **329**: p. 107022.

68. Hogben, H.J., M. Krzystyniak, G.T. Charnock, P.J. Hore, and I. Kuprov, *Spinach – a software library for simulation of spin dynamics in large spin systems.* Journal of Magnetic Resonance, 2011. **208**(2): p. 179-194.

69. Fokker, A.D., *Die mittlere Energie rotierender elektrischer Dipole im Strahlungsfeld.* Annalen der Physik, 1914. **348**(5): p. 810-820.

70. Planck, M., *Über einen Satz der statistischen Dynamik und seine Erweiterung in der Quantentheorie.* Sitzungsberichte der Königlich Preussischen Akademie der Wissenschaften zu Berlin, 1917: p. 324-341.

71. Kuprov, I., *Fokker-Planck formalism in magnetic resonance simulations.* Journal of Magnetic Resonance, 2016. **270**: p. 124-135.

72. Kuprov, I., *Other degrees of freedom*, in *Spin: from basic symmetries to quantum optimal control*. 2023, Springer. p. 181-221.

73. Allami, A.J., M.G. Concilio, P. Lally, and I. Kuprov, *Quantum mechanical MRI simulations: solving the matrix dimension problem.* Science Advances, 2019. **5**(7): p. eaaw8962.

74. Eills, J., W. Hale, and M. Utz, *Synergies between Hyperpolarized NMR and Microfluidics: a review.* Progress in Nuclear Magnetic Resonance Spectroscopy, 2022. **128**: p. 44-69.

75. Jones, J.A., *NMR quantum computation.* Progress in Nuclear Magnetic Resonance Spectroscopy, 2001. **38**(4): p. 325-360.

76. Cory, D.G., R. Laflamme, E. Knill, L. Viola, T.F. Havel, N. Boulant, G. Boutis, E. Fortunato, S. Lloyd, R. Martinez, C. Negrevergne, M. Pravia, Y. Sharf, G. Teklemariam, Y.S. Weinstein, and W.H. Zurek, *NMR Based Quantum Information Processing: Achievements and Prospects.* Fortschritte der Physik, 2000. **48**(9-11): p. 875-907.

77. Kuprov, I., *Mathematical Background*, in *Spin: From Basic Symmetries to Quantum Optimal Control*. 2023, Springer. p. 1-41.

78. Fick, A., *Über diffusion.* Annalen der Physik, 1855. **170**(1): p. 59-86.

79. Euler, L., *Principes généraux du mouvement des fluides.* Mémoires de l'académie des sciences de Berlin, 1757: p. 274-315.

80. Martin Jr, R. and M. Pierre, *Nonlinear reaction-diffusion systems*, in *Mathematics in science and engineering*. 1992, Elsevier. p. 363-398.

81. Goodwin, M.J., J.C. Dickenson, A. Ripak, A.M. Deetz, J.S. McCarthy, G.J. Meyer, and L. Troian-Gautier, *Factors that Impact Photochemical Cage Escape Yields.* Chemical Reviews, 2024. **124**(11): p. 7379-7464.

82. Khudyakov, I., A.A. Zharikov, and A.I. Burshtein, *Cage effect dynamics.* The Journal of Chemical Physics, 2010. **132**(1).

83. Lorand, J.P., *The cage effect.* Inorganic Reaction Mechanisms, Part II; Progress in Inorganic Chemistry, 1972: p. 207-325.

84. Bain, A.D. and J.S. Martin, *FT NMR of nonequilibrium states of complex spin systems, Part I: a Liouville space description.* Journal of Magnetic Resonance, 1978. **29**(1): p. 125-135.

85. Kuprov, I., *Incomplete Basis Sets*, in *Spin: From Basic Symmetries to Quantum Optimal Control*. 2023, Springer. p. 291-312.

86. Shishmarev, D., C.Q. Fontenelle, B. Linclau, I. Kuprov, and P.W. Kuchel, *Quantitative analysis of 2D EXSY NMR spectra of strongly coupled spin systems in transmembrane exchange.* ChemBioChem, 2024. **25**(3): p. e202300597.

87. Blumich, B., *NMR imaging of materials*. Vol. 57. 2000: OUP Oxford.

88. Price, W.S., *NMR studies of translational motion: principles and applications*. 2009: Cambridge University Press.

89. Valiullin, R., *Diffusion NMR of Confined Systems: Fluid Transport in Porous Solids and Heterogeneous Materials*. 2016: Royal Society of Chemistry.

90. Levante, T. and R. Ernst, *Homogeneous versus inhomogeneous quantum-mechanical master equations.* Chemical physics letters, 1995. **241**(1-2): p. 73-78.

91. Budd, C.J. and A. Iserles, *Geometric integration: numerical solution of differential equations on manifolds.* Philosophical Transactions of the Royal Society of London. Series A: Mathematical, Physical and Engineering Sciences, 1999. **357**(1754): p. 945-956.

92. Iserles, A., H.Z. Munthe-Kaas, S.P. Nørsett, and A. Zanna, *Lie-group methods.* Acta numerica, 2000. **9**: p. 215-365.

93. Casas, F. and A. Iserles, *Explicit Magnus expansions for nonlinear equations.* Journal of Physics A: Mathematical and General, 2006. **39**(19): p. 5445.

94. Blanes, S., F. Casas, and M. Thalhammer, *High-order commutator-free quasi-Magnus exponential integrators for non-autonomous linear evolution equations.* Computer Physics Communications, 2017. **220**: p. 243-262.

95. Rasulov, U., A. Acharya, M. Carravetta, G. Mathies, and I. Kuprov, *Simulation and design of shaped pulses beyond the piecewise-constant approximation.* Journal of Magnetic Resonance, 2023. **353**: p. 107478.

96. *Matlab R2024a*, in *The MathWorks, Natick, MA*. 2024.

97. McDonald, P.W., *The computation of transonic flow through two-dimensional gas turbine cascades*. Vol. 79825. 1971: American Society of Mechanical Engineers.

98. Versteeg, H.K. and W. Malalasekera, *An Introduction to Computational Fluid Dynamics: The Finite Volume Method*. 2007: Pearson Education Limited.

99. LeVeque, R.J., *Finite Volume Methods for Hyperbolic Problems*. 2002: Cambridge University Press.

100. Gauss, C.F., *Theoria attractionis corporum sphaeroidicorum ellipticorum homogeneorum methodo nova tractata*. 1813: Königliche Gesellschaft der Wissenschaften.

101. Voronoi, G., *Nouvelles applications des paramètres continus à la théorie des formes quadratiques. Premier mémoire. Sur quelques propriétés des formes quadratiques positives parfaites.* Journal für die Reine und Angewandte Mathematik, 1908. **1908**(133): p. 97-102.

102. Voronoi, G., *Nouvelles applications des paramètres continus à la théorie des formes quadratiques. Deuxième mémoire. Recherches sur les parallélloèdres primitifs.* Journal für die Reine und Angewandte Mathematik, 1908. **1908**(134): p. 198-287.

103. Kuprov, I., *Large-scale NMR simulations in liquid state: a tutorial.* Magnetic Resonance in Chemistry, 2018. **56**(6): p. 415-437.

104. Runge, C., *Über die numerische Auflösung von Differentialgleichungen.* Mathematische Annalen, 1895. **46**(2): p. 167-178.

105. Kutta, W., *Beitrag zur naherungsweisen Integration totaler Differentialgleichungen.* Zeitschrift für Mathematik und Physik, 1901. **46**: p. 435-453.

106. Munthe-Kaas, H., *Runge-Kutta methods on Lie groups.* BIT Numerical Mathematics, 1998. **38**(1): p. 92-111.

107. Fernandes, P., B. Plateau, and W.J. Stewart, *Efficient descriptor-vector multiplications in stochastic automata networks.* Journal of the ACM (JACM), 1998. **45**(3): p. 381-414.

108. Kuprov, I., *Coherent Spin Dynamics*, in *Spin: From Basic Symmetries to Quantum Optimal Control*. 2023, Springer. p. 107-179.

109. Kuprov, I., *Notes on Software Engineering*, in *Spin: From Basic Symmetries to Quantum Optimal Control*. 2023, Springer. p. 351-373.

110. Diels, O. and K. Alder, *Synthesen in der hydroaromatischen Reihe.* Justus Liebigs Annalen der Chemie, 1928. **460**(1): p. 98-122.

111. Arndt, T., P.K. Wagner, J.J. Koenig, and M. Breugst, *Iodine-Catalyzed Diels-Alder Reactions.* ChemCatChem, 2021. **13**(12): p. 2922-2930.

112. Lie, S., *Theorie der Transformationsgruppen*. 1888: BG Teubner Verlag.

113. Hinshelwood, C.N., *The kinetics of chemical change*. 1940: Oxford University, London.

114. Kuprov, I., *Diagonalization-free implementation of spin relaxation theory for large spin systems.* Journal of Magnetic Resonance, 2011. **209**(1): p. 31-38.

115. Goodwin, D.L. and I. Kuprov, *Auxiliary matrix formalism for interaction representation transformations, optimal control, and spin relaxation theories.* The Journal of chemical physics, 2015. **143**(8).

116. Peterson, K.A. and T.H. Dunning Jr, *Accurate correlation consistent basis sets for molecular core–valence correlation effects: The second row atoms Al–Ar, and the first row atoms B–Ne revisited.* The Journal of chemical physics, 2002. **117**(23): p. 10548-10560.

117. Zhao, Y. and D.G. Truhlar, *A density functional that accounts for medium-range correlation energies in organic chemistry.* Organic Letters, 2006. **8**(25): p. 5753-5755.

118. Frisch, M.J., G.W. Trucks, H.B. Schlegel, G.E. Scuseria, M.A. Robb, J.R. Cheeseman, G. Scalmani, V. Barone, B. Mennucci, G.A. Petersson, H. Nakatsuji, M. Caricato, X. Li, H.P. Hratchian, A.F. Izmaylov, J. Bloino, G. Zheng, J.L. Sonnenberg, M. Hada, M. Ehara, K. Toyota, R. Fukuda, J. Hasegawa, M. Ishida, T. Nakajima, Y. Honda, O. Kitao, H. Nakai, T. Vreven, J.A. Montgomery Jr., J.E. Peralta, F. Ogliaro, M.J. Bearpark, J. Heyd, E.N. Brothers, K.N. Kudin, V.N. Staroverov, R. Kobayashi, J. Normand, K. Raghavachari, A.P. Rendell, J.C. Burant, S.S. Iyengar, J. Tomasi, M. Cossi, N. Rega, N.J. Millam, M. Klene, J.E. Knox, J.B. Cross, V. Bakken, C. Adamo, J. Jaramillo, R. Gomperts, R.E. Stratmann, O. Yazyev, A.J. Austin, R. Cammi, C. Pomelli, J.W. Ochterski, R.L. Martin, K. Morokuma, V.G. Zakrzewski, G.A. Voth, P. Salvador, J.J. Dannenberg, S. Dapprich, A.D. Daniels, Ö. Farkas, J.B. Foresman, J.V. Ortiz, J. Cioslowski, and D.J. Fox, *Gaussian 09*. 2009, Gaussian, Inc.: Wallingford, CT, USA.

119. Marenich, A.V., C.J. Cramer, and D.G. Truhlar, *Universal Solvation Model Based on Solute Electron Density and on a Continuum Model of the Solvent Defined by the Bulk Dielectric Constant and Atomic Surface Tensions.* The Journal of Physical Chemistry B, 2009. **113**(18): p. 6378-6396.

120. London, F., *Théorie quantique des courants interatomiques dans les combinaisons aromatiques.* J. phys. radium, 1937. **8**(10): p. 397-409.

121. Le Bihan, D., *Looking into the functional architecture of the brain with diffusion MRI.* Nature reviews neuroscience, 2003. **4**(6): p. 469-480.

122. Mishra, R., A. Marchand, C. Jacquemmoz, and J.-N. Dumez, *Ultrafast diffusion-based unmixing of 1 H NMR spectra.* Chemical Communications, 2021. **57**(19): p. 2384-2387.

123. Mishra, R. and J.-N. Dumez, *Theoretical analysis of flow effects in spatially encoded diffusion NMR.* The Journal of Chemical Physics, 2023. **158**(1).

124. Redfield, A.G., *On the theory of relaxation processes.* IBM Journal of Research and Development, 1957. **1**(1): p. 19-31.

125. Moro, G. and J.H. Freed, *Calculation of ESR spectra and related Fokker–Planck forms by the use of the Lanczos algorithm.* The Journal of Chemical Physics, 1981. **74**(7): p. 3757-3773.

126. Torrey, H.C., *Nuclear spin relaxation by translational diffusion.* Physical Review, 1953. **92**(4): p. 962.

127. Pileio, G. and S. Ostrowska, *Accessing the long-time limit in diffusion NMR: The case of singlet assisted diffusive diffraction q-space.* Journal of Magnetic Resonance, 2017. **285**: p. 1-7.

128. Guduff, L., A.J. Allami, C. Van Heijenoort, J.-N. Dumez, and I. Kuprov, *Efficient simulation of ultrafast magnetic resonance experiments.* Physical Chemistry Chemical Physics, 2017. **19**(27): p. 17577-17586.

129. Guduff, L., I. Kuprov, C. Van Heijenoort, and J.-N. Dumez, *Spatially encoded 2D and 3D diffusion-ordered NMR spectroscopy.* Chemical Communications, 2017. **53**(4): p. 701-704.

130. Hwang, T.-L. and A. Shaka, *Water suppression that works: excitation sculpting using arbitrary wave-forms and pulsed-field gradients.* Journal of Magnetic Resonance, 1995. **112**(2): p. 275-279.

131. Trefethen, L.N., *Spectral methods in MATLAB*. 2000: SIAM.

132. Dumont, R.S., S. Jain, and A. Bain, *Simulation of many-spin system dynamics via sparse matrix methodology.* The Journal of chemical physics, 1997. **106**(14): p. 5928-5936.

133. Waudby, C.A. and J. Christodoulou, *GPU accelerated Monte Carlo simulation of pulsed-field gradient NMR experiments.* Journal of Magnetic Resonance, 2011. **211**(1): p. 67-73.

134. Kose, R. and K. Kose, *BlochSolver: A GPU-optimized fast 3D MRI simulator for experimentally compatible pulse sequences.* Journal of Magnetic Resonance, 2017. **281**: p. 51-65.

135. Maciejewski, M.W., A.D. Schuyler, M.R. Gryk, I.I. Moraru, P.R. Romero, E.L. Ulrich, H.R. Eghbalnia, M. Livny, F. Delaglio, and J.C. Hoch, *NMRbox: a resource for biomolecular NMR computation.* Biophysical journal, 2017. **112**(8): p. 1529-1534.